\documentclass[twocolumn,preprintnumbers,superscriptaddress,nofootinbib,aps,prd,floatfix,10pt]{revtex4-1}

\usepackage{amsmath,amssymb}
\usepackage{dsfont,cancel}
\usepackage{graphicx} 
\usepackage{epstopdf} 
\usepackage{slashed}
\usepackage{subfigure}
\usepackage{color}
\usepackage{multirow}
\usepackage[scientific-notation=true]{siunitx}
\usepackage{hyperref}
\usepackage[utf8]{inputenc}

\usepackage{footmisc}

\usepackage{array}
\newcolumntype{L}[1]{>{\raggedright\let\newline\\\arraybackslash\hspace{0pt}}m{#1}}
\newcolumntype{C}[1]{>{\centering\let\newline\\\arraybackslash\hspace{0pt}}m{#1}}
\newcolumntype{R}[1]{>{\raggedleft\let\newline\\\arraybackslash\hspace{0pt}}m{#1}}

\newcommand{\be}{\begin{eqnarray*}}
\newcommand{\ee}{\end{eqnarray*}}

\newcommand{\bee}{\begin{eqnarray}}
\newcommand{\eee}{\end{eqnarray}}
\newcommand{\beeq}{\begin{equation}}
\newcommand{\eeeq}{\end{equation}}

\newcommand{\BR}{{\text{BR}}}

\newcommand{\ttbar}{t \bar{t}}
\newcommand{\ttbb}{t \bar{t} b \bar{b}}
\newcommand{\ttZ}{t \bar{t} Z}
\newcommand{\ttX}{t \bar{t} X}
\newcommand{\ttS}{t \bar{t} S}
\newcommand{\ttA}{t \bar{t} A}
\newcommand{\ttZpV}{t \bar{t} Z'_\mathrm{V}}
\newcommand{\ttZpA}{t \bar{t} Z'_\mathrm{A}}
\newcommand{\ZpV}{Z'_\mathrm{V}}
\newcommand{\ZpA}{Z'_\mathrm{A}}

\newcommand{\pT}{p_\mathrm{T}}
\newcommand{\pTX}{p_{\mathrm{T},X}}
\newcommand{\CLs}{\mathrm{CL}_\mathrm{s}}

\definecolor{colorMD}{rgb}{0.0,0.8,0.2}

\begin{document}

\title{Determining the Quantum Numbers of Simplified Models in $t\bar{t}X$ production at the LHC}

\author{Matthew J. Dolan} 
\affiliation{ARC Centre of Excellence for Particle Physics at the Terascale,\\
  School of Physics, The University of Melbourne, Victoria 3010, Australia\\[0.1cm]}

\author{Michael Spannowsky} 
\affiliation{Institute for Particle Physics Phenomenology, Department
  of Physics,\\Durham University, DH1 3LE, United Kingdom\\[0.1cm]}
\author{Qi Wang} 
\affiliation{Institute for Particle Physics Phenomenology, Department
  of Physics,\\Durham University, DH1 3LE, United Kingdom\\[0.1cm]}
\author{Zhao-Huan Yu}
\affiliation{ARC Centre of Excellence for Particle Physics at the Terascale,\\
  School of Physics, The University of Melbourne, Victoria 3010, Australia\\[0.1cm]}

\begin{abstract}
Simplified models provide an avenue for characterising and exploring New Physics for large classes of UV theories. In this article we study the ability of the LHC to probe the spin and parity quantum numbers of a new light resonance $X$ which couples predominantly to the third generation quarks in a variety of simplified models through the $t\bar t  X$ channel. After evaluating the LHC discovery potential for $X$, we suggest several kinematic variables sensitive to the spin and CP properties of the new resonance. We show how an analysis exploiting differential distributions in the semi-leptonic channel can discriminate among various possibilities. We find that the potential to discriminate a scalar from a pseudoscalar or (axial) vector to be particularly promising.
\end{abstract}

\pacs{}
\preprint{IPPP/16/47, DCPT/16/94}

\maketitle

\section{Introduction}

After the discovery of the Higgs boson \cite{Aad:2012tfa,Chatrchyan:2012xdj}, the particle physics community is eagerly expecting and awaiting the discovery of Beyond-the-Standard-Model (BSM) physics at the Large Hadron Collider. While the majority of attention from phenomenologists has been focussed on the possibility of heavy new particles at the high energy frontier, it could also be the case that light resonances have escaped notice from previous colliders such as LEP and the Tevatron, and may be discovered in the large datasets which the LHC will accrue in coming years.

This has been an area of particular interest due to the Galactic Centre excess of diffuse gamma-rays~\cite{Hooper:2010mq,Gordon:2013vta,Daylan:2014rsa,Calore:2014xka}, which may be explained by dark matter (DM) annihilating via a light mediator into Standard Model (SM) particles~\cite{Agrawal:2014oha,Calore:2014nla,Daylan:2014rsa}. In these cases the excess can be explained with scalar~\cite{Boehm:2014bia,Ko:2014gha,Abdullah:2014lla,Yu:2014mfa},  pseudoscalar~\cite{Boehm:2014hva}, or axial (or mixed coupling) vector mediators~\cite{Berlin:2014tja,Hooper:2014fda}. The collider sensitivity to these mediators have been explored in a series of papers under the assumption that the mediator couples to the quarks and DM~\cite{Buchmueller:2013dya,Buchmueller:2014yoa,Malik:2014ggr,Buckley:2014fba,Harris:2014hga,Xiang:2015lfa,Haisch:2015ioa,Boveia:2016mrp}. Since some analyses of the GC excess suggest that DM annihilation into $b$ quarks provides a particularly good fit, some studies have assumed the mediator predominantly couples to the third generation quarks. In that case an important collider signal is associated production with a pair of top or bottom quarks~\cite{Kozaczuk:2015bea,Craig:2015jba,Casolino:2015cza}, particularly for spin-1 mediators where LHC production via gluon fusion is forbidden by the Landau-Yang theorem~\cite{Landau:1948kw,Yang:1950rg}. There has also been model building interest recently in top-philic $Z'$ bosons in the context of a slight excess in $t\bar{t}h$ searches for SM Higgs boson production~\cite{Cox:2015afa}. Some recent work has studied searches for $\ttbar$ resonances in the context of two Higgs doublet models~\cite{Chen:2015fca,Gori:2016zto,Craig:2016ygr}, and on searches for top-philic dark matter mediators~\cite{Arina:2016cqj}. However, these works focus on the heavier resonances. 
It is therefore important to understand the ability of the LHC to discover and measure the properties of light new resonances with strong couplings to the third generation. 

If such a new light (i.e. $m_X\lesssim 100$~GeV) resonance $X$ is discovered in Run 2 of the LHC, a first priority will be the characterisation of its quantum numbers. In the context of resonances with strong coupling to top quarks studies have already been made in $\ttX$ production~\cite{Buckley:2015vsa,Buckley:2015ctj,Casolino:2015cza}, focussing on the semi- and di-leptonic top decay channels, where either one or both tops decay leptonically. In the case of di-leptonic top decays, it is known (building upon older work on spin-polarisation in $\ttbar$ production~\cite{Mahlon:1995zn}) that the azimuthal angle between the leptons encodes much of the relevant CP information. Related work has focussed on dijet angular correlations in $pp\to jj X$~\cite{Haisch:2013fla}, as well as extending results to NLO accuracy~\cite{Demartin:2014fia,Backovic:2015soa}.

In this paper we seek to extend these previous works in a number of ways. We  explore other angular variables which may be of use in pinning down the quantum numbers of top-philic resonances at the LHC. Where most other works have focussed on the di-leptonic final state (with some exceptions~\cite{Ellis:2013yxa, Casolino:2015cza}) we perform a detector-level analysis of the semi-leptonic final states. We find that although the SM backgrounds are challenging, this final state will indeed be useful in the discovery and characterisation of new light resonances.

The structure of our paper is as follows: in Sec.~\ref{sec:SMS} we introduce and specify the simplified models which we will study, and in Sec.~\ref{sec:parton} we introduce at parton level a number of different kinematic distributions which exhibit some sensitivity to the quantum numbers of the new resonance, before in Sec.~\ref{sec:detector} we perform a full analysis in the semi-leptonic final state of the LHC sensitivity, providing both estimated discovery reach and cross sections required for the LHC to be able to discriminate among different simplified models. 

\section{Simplified Models}
\label{sec:SMS}

We study the phenomenology of a variety of simplified models~\cite{Alves:2011wf} with a new neutral resonance which we assume to be an eigenstate of parity and charge conjugation. Its couplings are restricted to the third generation quarks (bottoms and tops) only.
For a scalar resonance $S$ and a pseudoscalar resonance $A$, we assume the following CP-conserving interaction Lagrangians:
\begin{eqnarray}
  \mathcal{L}_{\rm{S}} &=&  - \sum_{q=b,t} \frac{g_q y_q}{\sqrt{2}} S \bar q q , \\
  \mathcal{L}_{\rm{P}} &=&  - \sum_{q=b,t} \frac{g_q y_q}{\sqrt{2}} A\bar q i\gamma_5 q,
\end{eqnarray}
where $y_q$ is the SM Yukawa couplings.
We also study a vector resonance $\ZpV$ and an axial vector resonance $\ZpA$ with interaction Lagrangians given by
\begin{eqnarray}
  \mathcal{L}_{\rm{V}} &=&  -  \sum_{q=b,t} g_q Z'^{\mu}_\mathrm{V} \bar{q}\gamma^{\mu}q ,\\
  \mathcal{L}_{\rm{AV}} &=&  -  \sum_{q=b,t} g_q Z'^{\mu}_\mathrm{A} \bar{q}\gamma_{\mu}\gamma_5 q.
\end{eqnarray}

In all these cases the decay width of the resonance is set to its natural width calculated from the theory parameters at tree level. We do not include any interactions between $X$ and possible dark matter candidates, focussing on its interactions with the SM (equivalently, there may be a coupling between $X$ and dark matter, but we study the parameter space where $m_X/2>m_{\rm{DM}}$). In the case that the resonance is lighter than $2m_t$ it must decay into a pair of $b$ quarks with a branching ratio equal to one (neglecting three-body decays).
While these Lagrangians will also lead to dimension-five interactions with gluons for the scalar and pseudoscalar (whose CP properties can be probed in dijet angular correlations for instance~\cite{Dolan:2014upa}), in this paper we exclusively focus on what can be gleaned from  associated production with tops.

We have implemented these models in \texttt{FeynRules}~\cite{Alloul:2013bka} which allows us to generate simulated events at the LHC using \texttt{MadGraph}~\cite{Alwall:2014hca} via the UFO~\cite{Degrande:2011ua} format.

\section{Spin and Parity Discriminating Variables}
\label{sec:parton}

\begin{figure}[t!]
  \centering
  \includegraphics[width=0.46\textwidth]{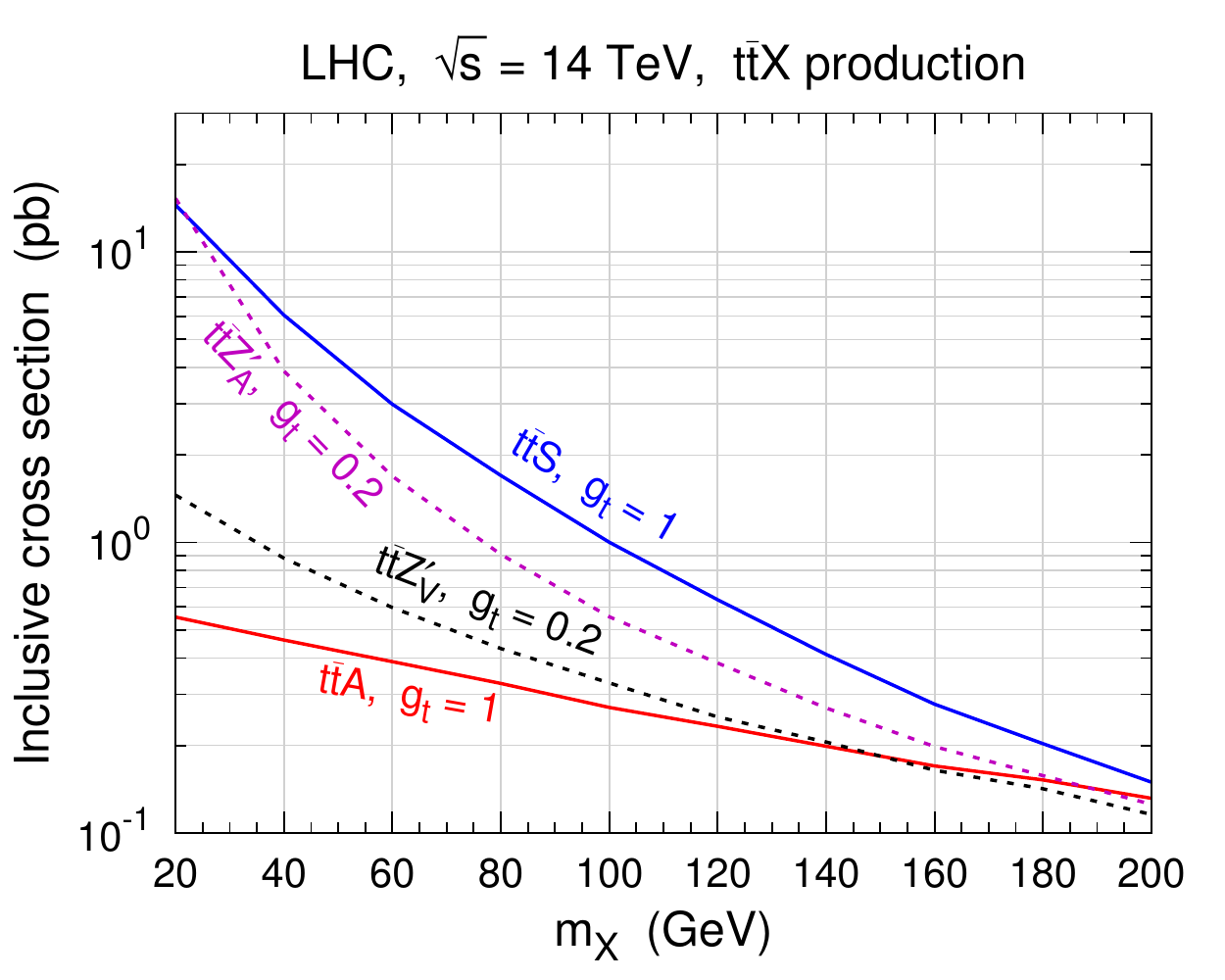}
  \caption{This figure shows the $\ttX +\text{jets}$ ($X=S$, $A$, $\ZpV$, or $\ZpA$) production cross sections as functions of $m_X$ at the 14~TeV LHC for the four simplified models.}
  \label{fig:mass_Xsec}
\end{figure}

In Fig.~\ref{fig:mass_Xsec} we show the behaviour of the cross-sections for the four simplified models as functions of the resonance mass $m_X$. As has been demonstrated before~\cite{Casolino:2015cza}, for low masses (below around 100~GeV) the $\ttA$ production cross section is quite suppressed relative to that of $\ttS$, and is smaller by over an order of magnitude below 40~GeV. We also observe similar behaviour (although not as extreme) in the $\ttZpV$ versus $\ttZpA$ cross sections. The differences between the cross sections become smaller as $m_X$ increases, and are all within a factor of two at $m_X=200$~GeV.

To attempt to understand this we have calculated the helicity amplitudes for $\ttbar A$ and $\ttbar S$ production, using the Weyl-van-der-Waerden spinous formalism for the case of massive particles~\cite{Dittmaier:1998nn}. However, due to the complicated 3-body phase space for the processes we study it is difficult to leverage these analytic results into an analytic understanding of the origin of the different cross-section values. As one might expect, the presence of the $\gamma^5$ appears to be responsible: comparing the scalar and pseudoscalar cases the $\gamma^5$ leads to sign change for a number of the helicity amplitudes, suggesting a destructive interference in the pseudoscalar scenario. 

We now turn to the information available in the kinematic distributions which can be formed from the $\ttX$ final state ($X=S$, $A$, $\ZpV$, $\ZpA$), focussing on those which have particular sensitivity to the CP and spin properties. For clarity, we present our results in this section at parton level, before providing a full detector-level analysis in Section~\ref{sec:detector}.

\begin{figure*}[t!]
  \centering
  \subfigure[~Normalised $m_{\ttbar}$ distributions\label{fig:parton:dist:a}]{
  \includegraphics[width=0.46\textwidth]{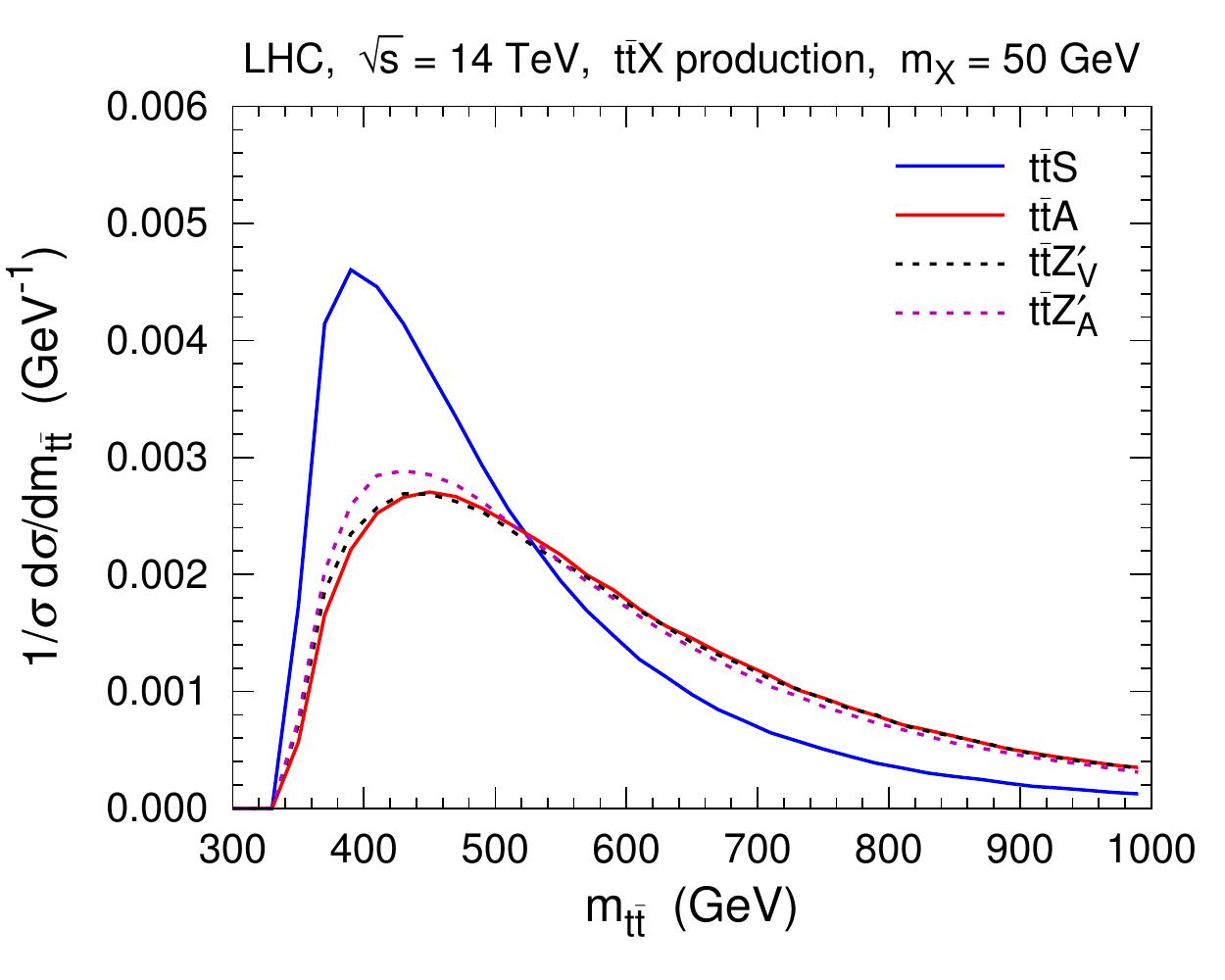}}
  \hfill
  \subfigure[~Normalised $\pTX$ distributions\label{fig:parton:dist:b}]{
  \includegraphics[width=0.46\textwidth]{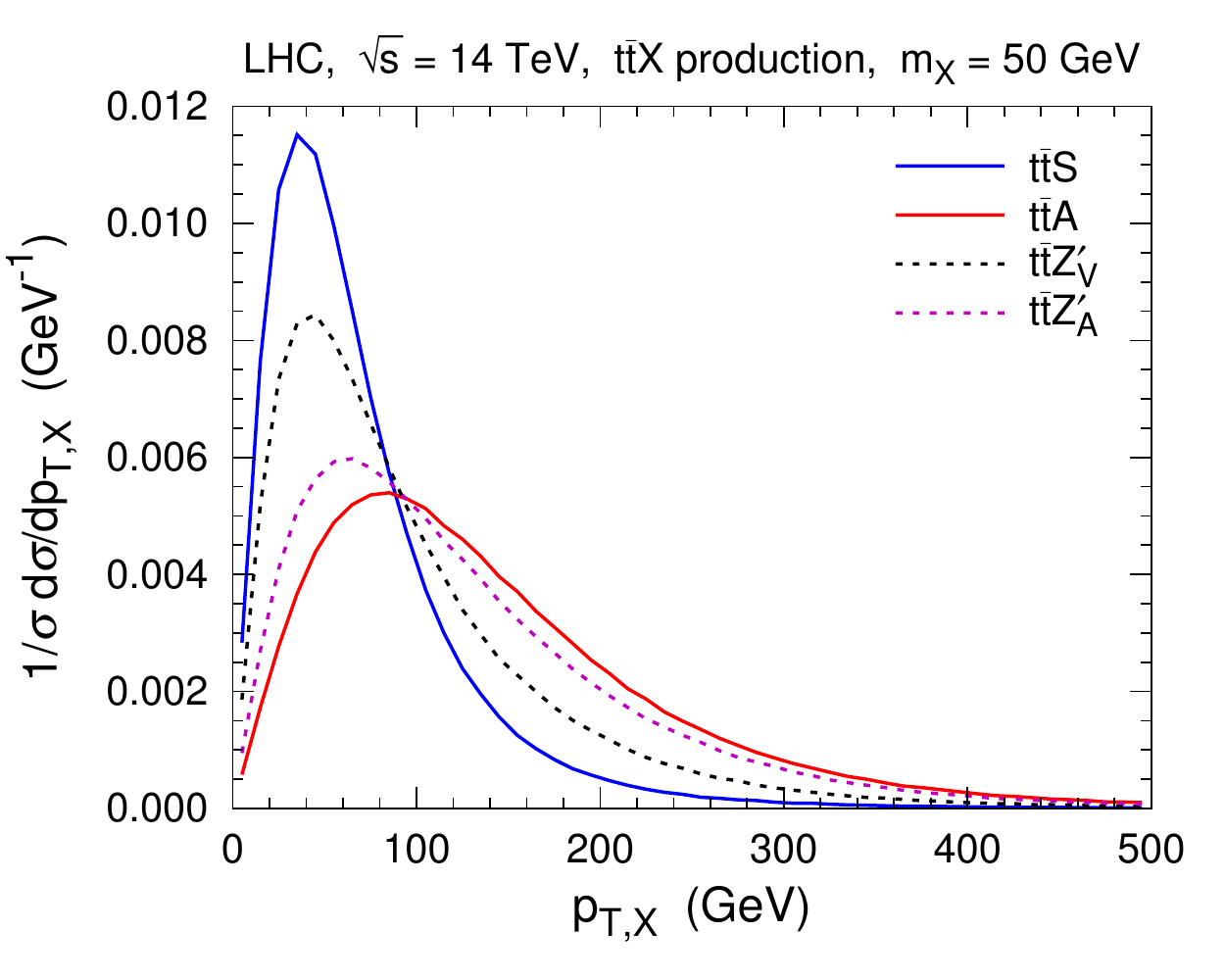}}
  \subfigure[~Normalised $\theta_t^\mathrm{CM}$ distributions\label{fig:parton:dist:c}]{
  \includegraphics[width=0.46\textwidth]{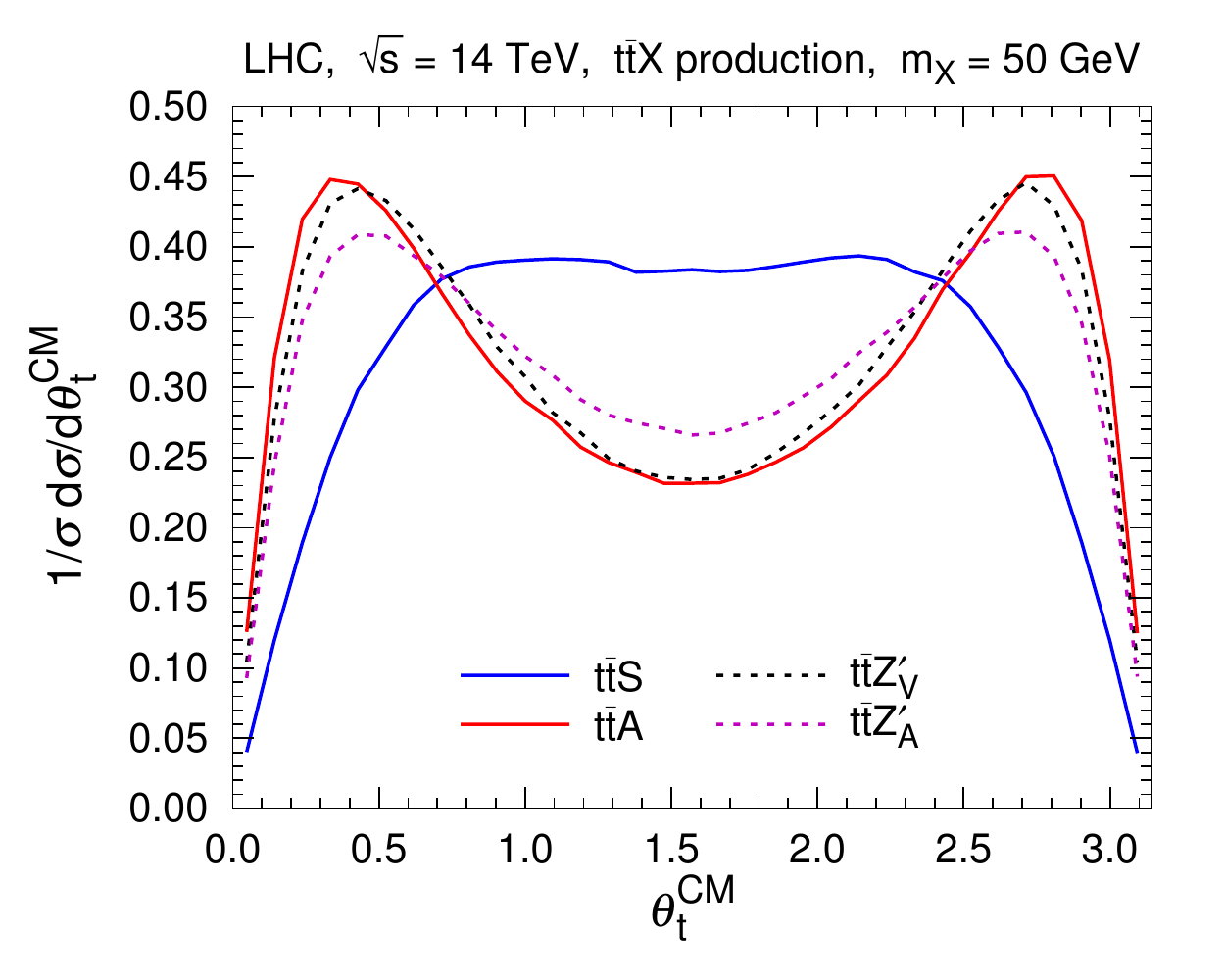}}
  \hfill
  \subfigure[~Normalised $\Theta^\mathrm{CM}$ distributions\label{fig:parton:dist:d}]{
  \includegraphics[width=0.46\textwidth]{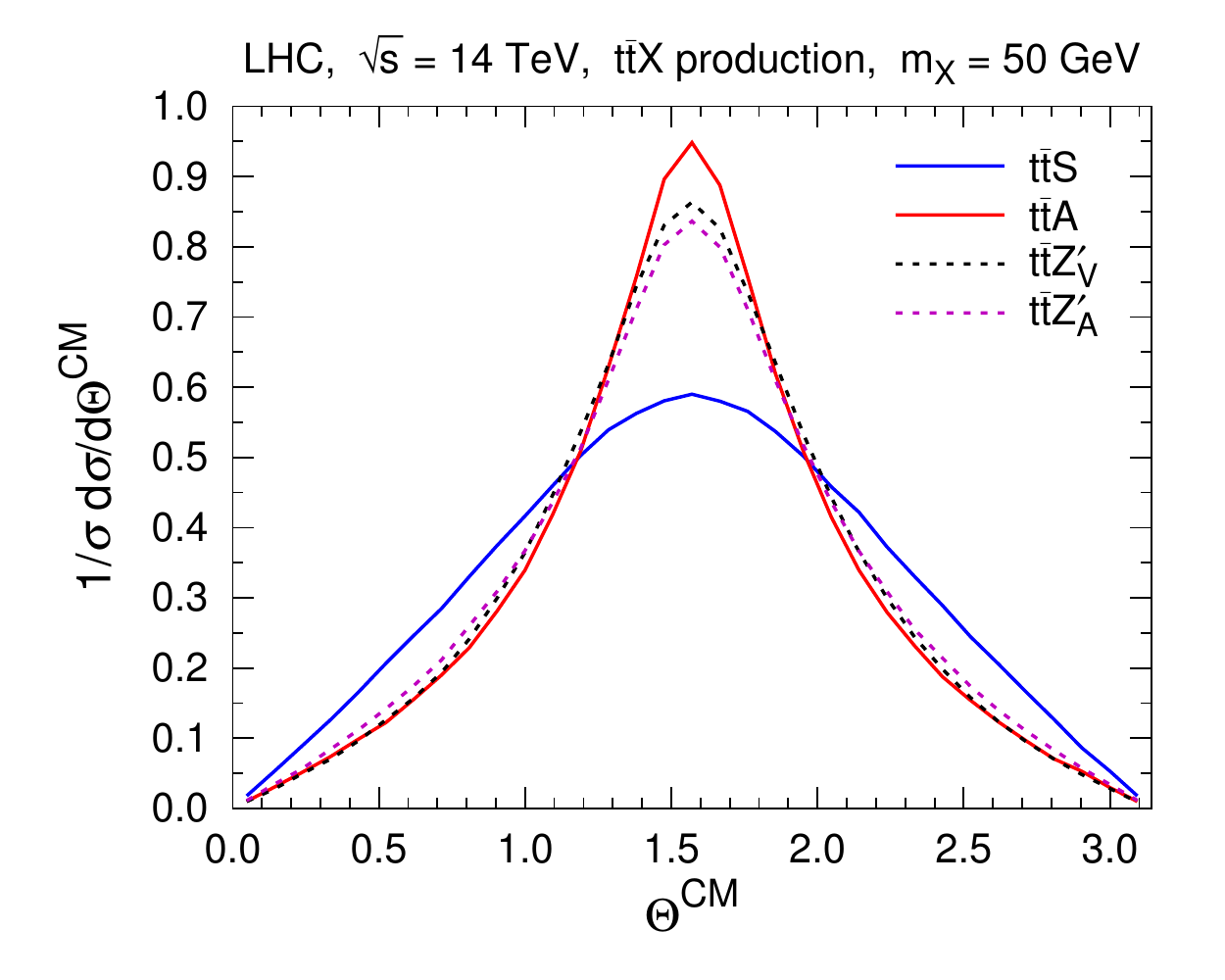}}
  \caption{This figure shows the normalised distributions of $m_{\ttbar}$ (a), $\pTX$ (b), $\theta_t^\mathrm{CM}$ (c) and $\Theta^\mathrm{CM}$ (d) for parton-level $\ttbar X$ production with $\sqrt{s}=14$~TeV and $m_X=50~\si{GeV}$. Here $X=S$, $A$, $\ZpV$, or $\ZpA$. The angular variables $\theta_t^\mathrm{CM}$ and $\Theta^\mathrm{CM}$ are defined in the CM frame of the $\ttbar X$ system. As shown in Fig.~\ref{fig:angles}, $\theta_t^\mathrm{CM}$ is the angle between $t$ and the beamline, while $\Theta^\mathrm{CM}$ is the angle between the normal vector to the $\ttbar X$ system and the beamline.}
  \label{fig:parton:dist}
\end{figure*}

We show in Figs.~\ref{fig:parton:dist:a} and \ref{fig:parton:dist:b} the distributions of the di-top invariant mass $m_{\ttbar}$ and the transverse momentum of the resonance $\pTX$, for the four simplified models introduced above with the benchmark mass of $m_X=50~\si{GeV}$. The distributions are normalised and hence independent of the coupling $g_t$, because we are primarily interested in the shape of the distributions rather than the precise values of the production cross sections.
Both $m_{\ttbar}$ and $\pTX$ (which are correlated) have previously been suggested as variables which may help distinguish between $\ttS$ and $\ttA$ production~\cite{Buckley:2015vsa,Buckley:2015ctj,Ellis:2013yxa}; here we see that these variables are also sensitive to $\ttZpV$ and $\ttZpA$ production. The distributions are generally quite similar in shape. However, we notice that $\ttA$ leads to the hardest distributions, with a shift in the peak and a longer tail at large $m_{\ttbar}$ and $\pTX$ compared to $\ttS$. $\ttZpV$ and $\ttZpA$ interpolate between these two behaviours: they lead to spectra which are harder than $\ttS$, but not so much as $\ttA$.

It is known that the azimuthal angle distribution between the two top quarks incorporates much information about the quantum numbers of the resonance $X$. Accessing this information is non-trivial however: the only case where the both tops could in principle be fully reconstructed without missing energy is the fully-hadronic scenario, which for any realistic analysis will be plagued by insurmountable QCD backgrounds. This has led Refs.~\cite{Buckley:2015vsa,Buckley:2015ctj} (based on previous work on $\ttbar$ spin correlations~\cite{Mahlon:1995zn,Mahlon:2010gw}) to explore the fully leptonic case, substituting the azimuthal angle between the leptons for the top quarks, an idea which has met with some success even when $X$ decays to dark matter~\cite{Buckley:2015ctj}. They have shown that constraints can be set on the top-quark Yukawa coupling in associated production towards the end of LHC Run 2 using these techniques.  

\begin{figure}[t!]
  \centering
  \includegraphics[width=0.35\textwidth]{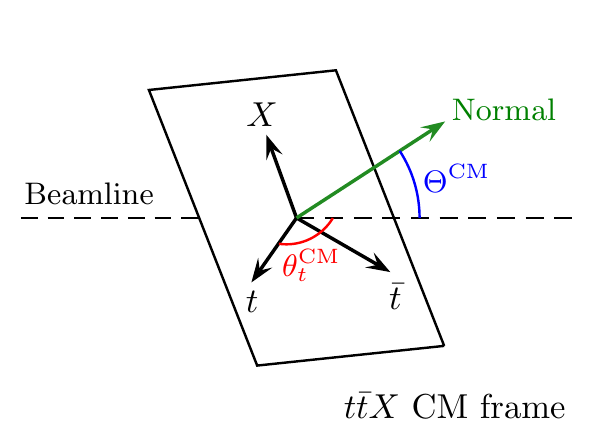}
  \caption{This figure shows the definitions of $\theta_{t}^\mathrm{CM}$ and $\Theta^\mathrm{CM}$ in the CM frame of the $\ttX$ system. $\theta_{t}^\mathrm{CM}$  is the angle between the $t$-quark and the beamline and $\Theta^\mathrm{CM}$ is the angle between the normal vector to the $\ttbar X$ system and the beamline. }
  \label{fig:angles}
\end{figure}

We therefore consider other angular variables, derived from boosting to the centre-of-mass (CM) frame of the reconstructed $\ttX$ system. We have investigated a variety of different constructions, and present results for two of the most sensitive that we have found. As illustrated in Fig.~\ref{fig:angles}, $\theta_{t}^\mathrm{CM}$ is the angle between $t$ and the beamline in the CM frame. The normalised $\theta_{t}^\mathrm{CM}$ distributions are shown in Fig.~\ref{fig:parton:dist:c}. We find that this variable is particularly sensitive to $\ttS$ production, which exhibits a broad plateau at $\pi/2$. The other processes all have a double-peak structure, with $\ttA$ being the sharpest defined, and $\ttZpV$ and $\ttZpA$ (similar to $m_{\ttbar}$ and $\pT$ above) interpolating between $\ttS$ and $\ttA$.

The other angular variable $\Theta^\mathrm{CM}$ utilises the fact that in the CM frame the $\ttX$ system forms a plane. We consider the normal vector to this plane, and $\Theta^\mathrm{CM}$ is angle between the normal and the beamline, as explained in Fig.~\ref{fig:angles}.
Fig.~\ref{fig:parton:dist:d} shows the normalised $\Theta^\mathrm{CM}$ distributions. The shape differences between the scalar and other resonances are not as great in this case, with the distributions for all the simplified models peaking at $\pi/2$. The $\ttS$ distribution is notable only in that it has the broadest distribution among them. While these variables show good sensitivity to the properties of the resonance $X$, a more realistic assessment of their utility requires a full analysis to be performed, which we now turn to.

\section{Detector-level Analysis}
\label{sec:detector}

In this section we perform a detector-level LHC analysis for $\sqrt{s}=14~\si{TeV}$ incorporating the dominant backgrounds and reconstruction effects using the variables from the previous section to evaluate resonance properties. We focus on the challenging semi-leptonic final state where the hadronic top can be fully reconstructed, but where jet backgrounds are a larger problem than the dileptonic case. Ultimately, due to the small cross sections involved in $\ttX$ production, utilising both the semi-leptonic and leptonic final states will lead to a better understanding on the class of models we consider.

\subsection{Event Simulation and Analysis Details}

Previous work has demonstrated that the dominant background in $\ttbar X$ final states for low resonance masses comes from $\ttbar b\bar b$ production~\cite{Casolino:2015cza,Moretti:2015vaa}. While the $\ttbar+\text{light jets}$ rate is significant it is a subdominant background after $b$-tagging, but with similar kinematics to $\ttbar b\bar b$ and so will be suppressed by the same analysis cuts. We also include $\ttZ$ which is more subdominant, but important for a possible data-driven background estimation. We generate background and signal samples with \texttt{MadGraph~5.2}~\cite{Alwall:2014hca} before showering them with \texttt{PYTHIA~6}~\cite{Sjostrand:2006za} and passing them through the \texttt{Delphes~3}~\cite{deFavereau:2013fsa} detector simulation using the default ATLAS detector card. Thus, for jet $\pT = 100~\si{GeV}$, the $b$-tagging efficiency is assumed to be 73\%, with the misidentification rates of $c$-jets and other light jets being 14\% and 0.27\%, respectively. Jets are clustered using the anti-$k_\mathrm{T}$ algorithm~\cite{Cacciari:2008gp} with an angular distance parameter $R = 0.4$.

\begin{figure}[t!]
  \centering
  \includegraphics[width=0.46\textwidth]{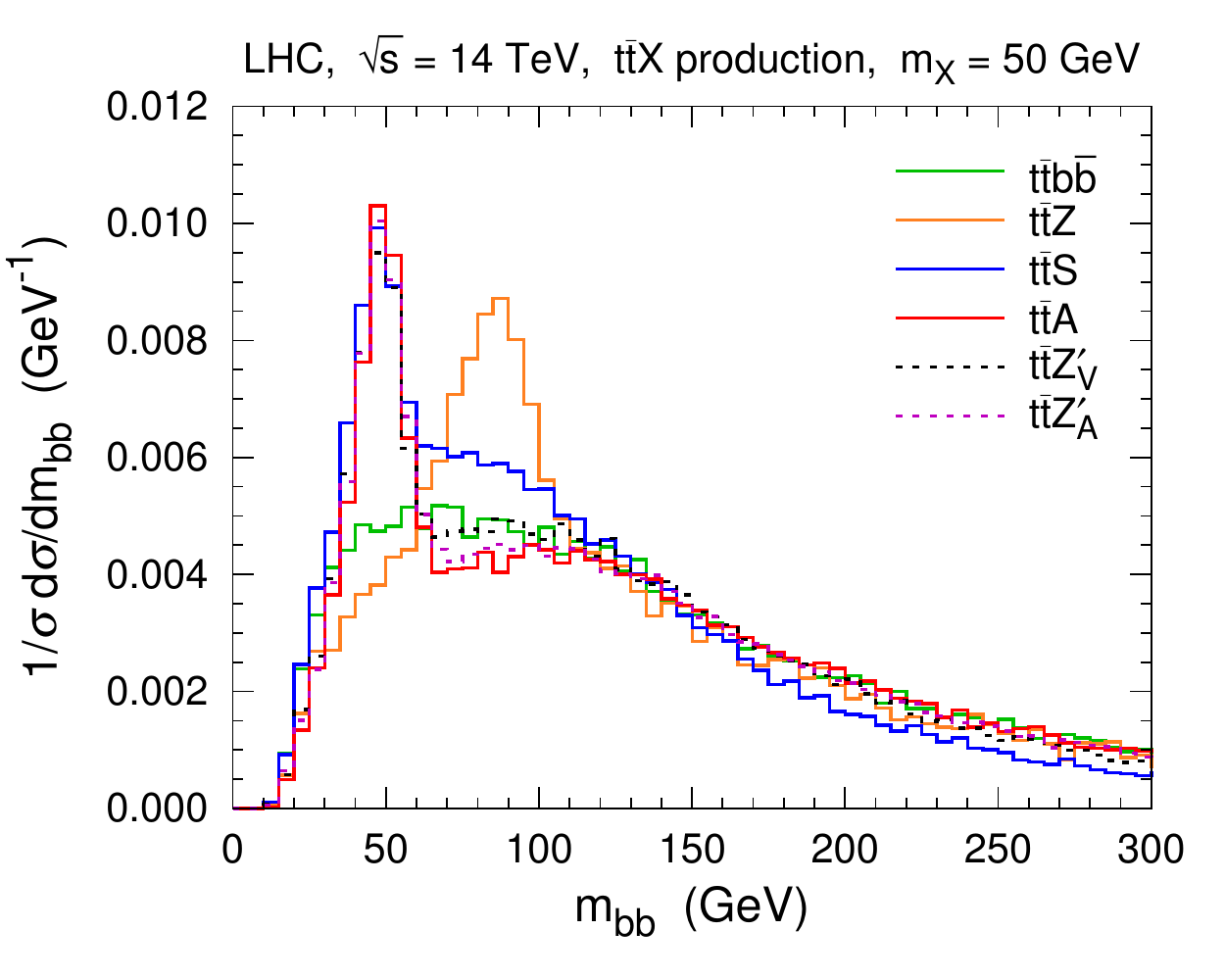}
  \caption{This figure shows the normalised $m_{bb}$ distributions after applying all the selection cuts except the cut on $m_{bb}$. For all the signals we take $m_X = 50~\si{GeV}$.}
  \label{fig:delphes:mbb}
\end{figure}

Selection cuts are adopted as follows. Firstly we require the selected events to contain exactly one charged lepton $\ell$ (electron or muon), exactly four $b$-tagged jets and at least two light jets. The lepton must be isolated from any jet via the condition  $\Delta R > 0.4$. Moreover, the lepton and the jets should have $\pT > 25~\si{Ge~V}$ and $|\eta| < 2.5$.

To reconstruct the hadronically decaying top we iterate through the reconstructed light jets and $b$-jets and find the combination which minimises
\begin{equation}
\chi^2 = \frac{(m_{jj}-m_W)^2}{m_W^2} + \frac{(m_{t,\mathrm{had}}-m_t)^2}{m_t^2}\, ,
\end{equation}
where $m_{jj}$ is the invariant mass of two light jets $j_1$ and $j_2$, and $m_{t,\mathrm{had}}$ is the invariant mass of $j_1$, $j_2$, and a $b$-jet $b_1$. After that, to reconstruct the leptonically decaying top we iterate through the remaining $b$-jets and find the one $b_2$ which minimises
\begin{equation}
\chi^2 = \frac{(m_{t,\mathrm{lep}}-m_t)^2}{m_t^2} \, ,
\end{equation}
where $m_{t,\mathrm{lep}}$ is the invariant mass constructed by $b_2$, the lepton, and the missing transverse momentum $\slashed{\mathbf{p}}_\mathrm{T}$.
The remaining $b$-jets $b_3$ and $b_4$ are used to search for the resonance $X$. We denote their invariant mass as $m_{bb}$ and show the normalised distributions in Fig.~\ref{fig:delphes:mbb} for our benchmark point with $m_X=50~\si{GeV}$. There is a clear peak at the signal resonance position, and the $\ttbb$ background is flat in the vicinity of the signal. As a reference for calibration, we also show the distribution of the $\ttZ$ background, which exhibits a clear $Z$ peak. We observe that all signals exhibit long tails in the $b\bar b$ invariant mass, due to misattribution of the $b$'s from the tops and the $b$'s from the resonance. Based on experience with the SM Higgs, the use of boosted techniques should ameliorate this.

\begin{table}[!t]
\centering
\setlength\tabcolsep{.4em}
\renewcommand{\arraystretch}{1.3}
\caption{Expected background and signal events per \si{fb^{-1}} after each step of the selection cuts for $m_X=50~\si{GeV}$. We take $g_q = 1$  $\ttS$ and $\ttA$ signals, and $g_q = 0.2$ is assumed for the $\ttZpV$ and $\ttZpA$ signals.}
\label{tab:cutflow}
\begin{tabular}{cccccccccc}
\hline\hline
 & $\ttbb$ & $\ttS$ & $\ttA$ & $\ttZpV$ & $\ttZpA$ \\
\hline
No cut                                      & 24375 & 4211 & 428   & 714   & 241  \\
1 lepton                                    & 4612  & 744  & 80.0  & 132   & 44.4 \\
4 $b$-tags                                  & 106   & 33.9 & 5.15  & 7.12  & 27.5 \\
$\geq$ 2 light jets                         & 72.9  & 22.1 & 3.51  & 4.86  & 18.7 \\
$m_{jj} \in (60,100)~\si{GeV}$              & 42.0  & 12.6 & 2.05  & 2.82  & 10.9 \\
$m_{t,\mathrm{had}} \in (120,200)~\si{GeV}$ & 39.1  & 11.9 & 1.92  & 2.64  & 10.2 \\
$m_{t,\mathrm{lep}} \in (120,220)~\si{GeV}$ & 30.2  & 9.87 & 1.52  & 2.09  & 8.07 \\
$m_{bb} \in (35,65)~\si{GeV}$               & 4.35  & 2.33 & 0.333 & 0.450 & 1.78 \\
\hline\hline
\end{tabular}
\end{table}

To further isolate the signal we impose the selection cuts $60~\si{GeV}<m_{jj}<100~\si{GeV}$, $120~\si{GeV}<m_{t,\mathrm{had}}<200~\si{GeV}$, $120~\si{GeV}<m_{t,\mathrm{lep}}<220~\si{GeV}$, and $35~\si{GeV}<m_{bb}<65~\si{GeV}$. The expected yields per inverse femtobarn for the $\ttbb$ background and the signals after each steps of selection cuts are presented in Table~\ref{tab:cutflow}. These cuts suppress the $\ttbb$ background by a factor of $\sim\num{5e3}$.
The $\ttZ$ background is lower than $\ttbb$ by two orders of magnitude.

\begin{figure*}[t!]
  \centering
  \subfigure[~$\ttS$ production]{\includegraphics[width=0.46\textwidth]{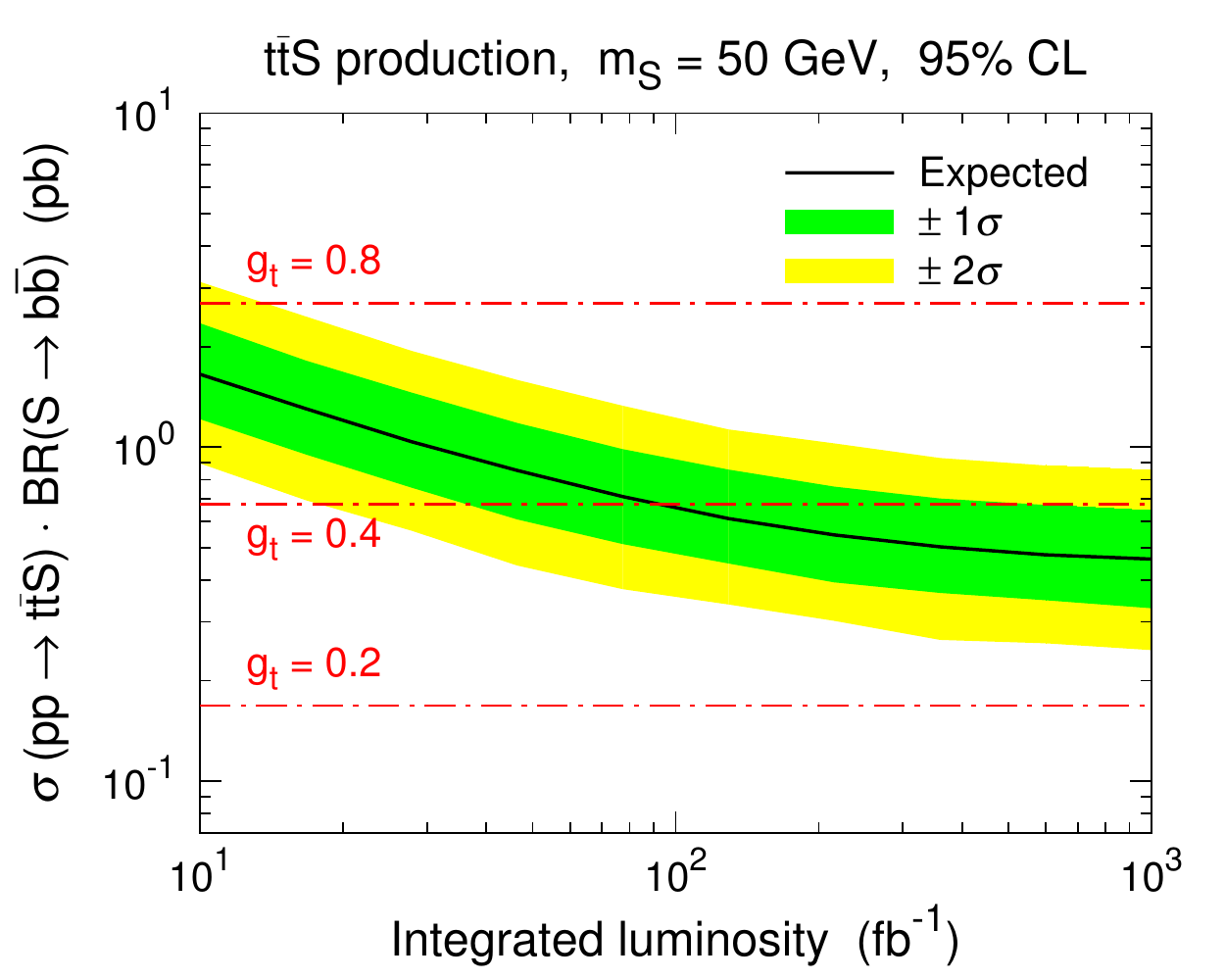}}
  \hfill
  \subfigure[~$\ttA$ production]{\includegraphics[width=0.46\textwidth]{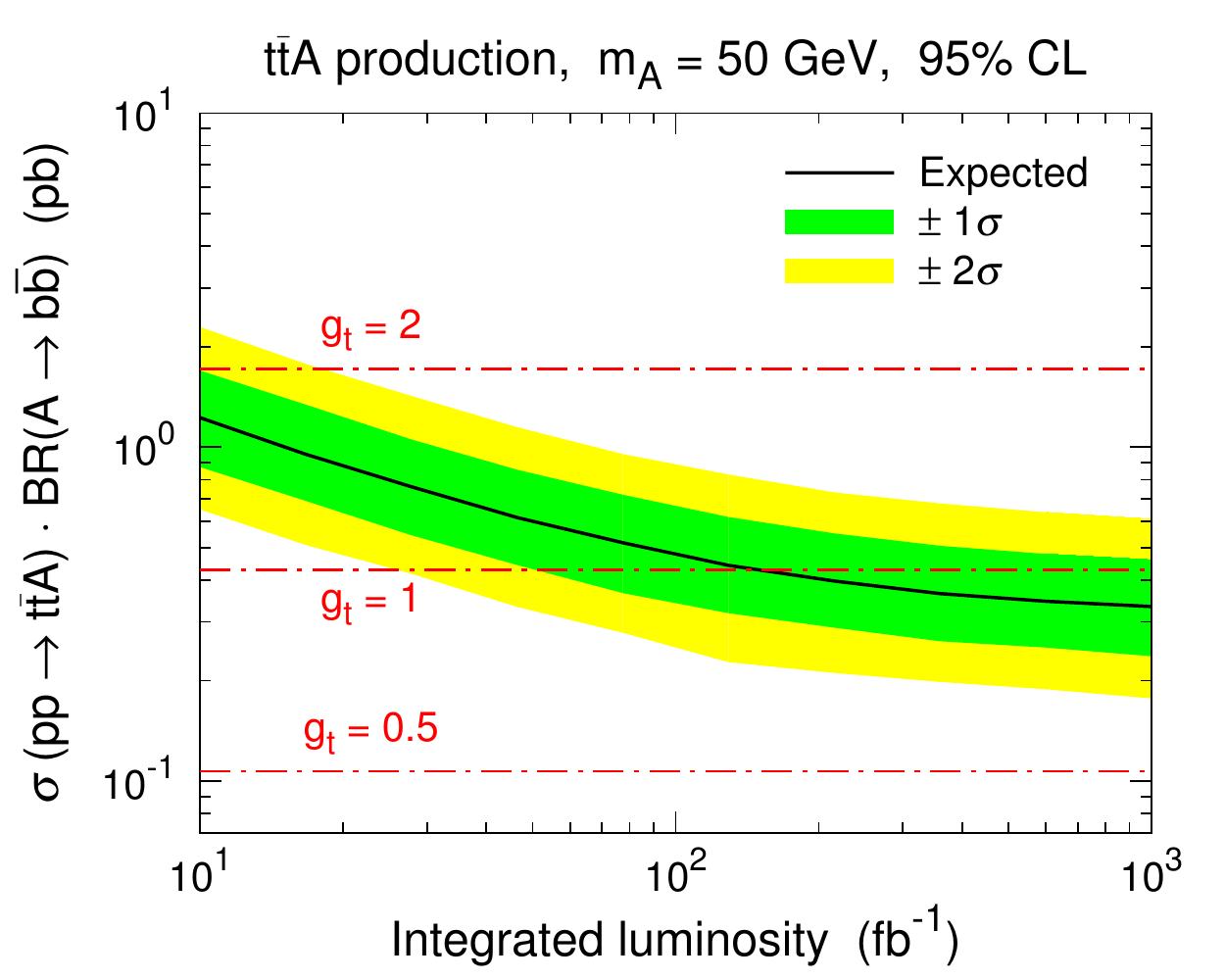}}
  \subfigure[~$\ttZpV$ production]{\includegraphics[width=0.46\textwidth]{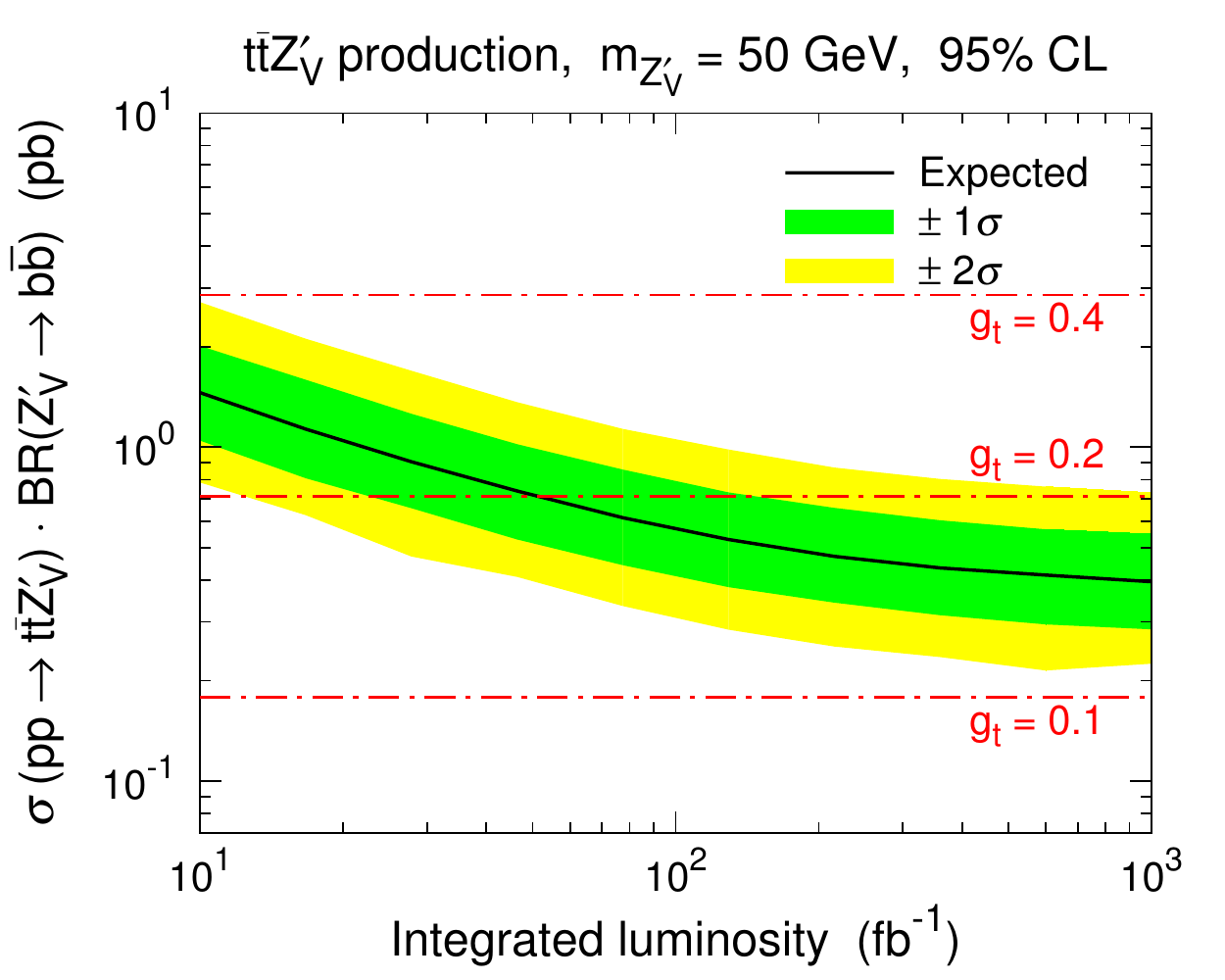}}
  \hfill
  \subfigure[~$\ttZpA$ production]{\includegraphics[width=0.46\textwidth]{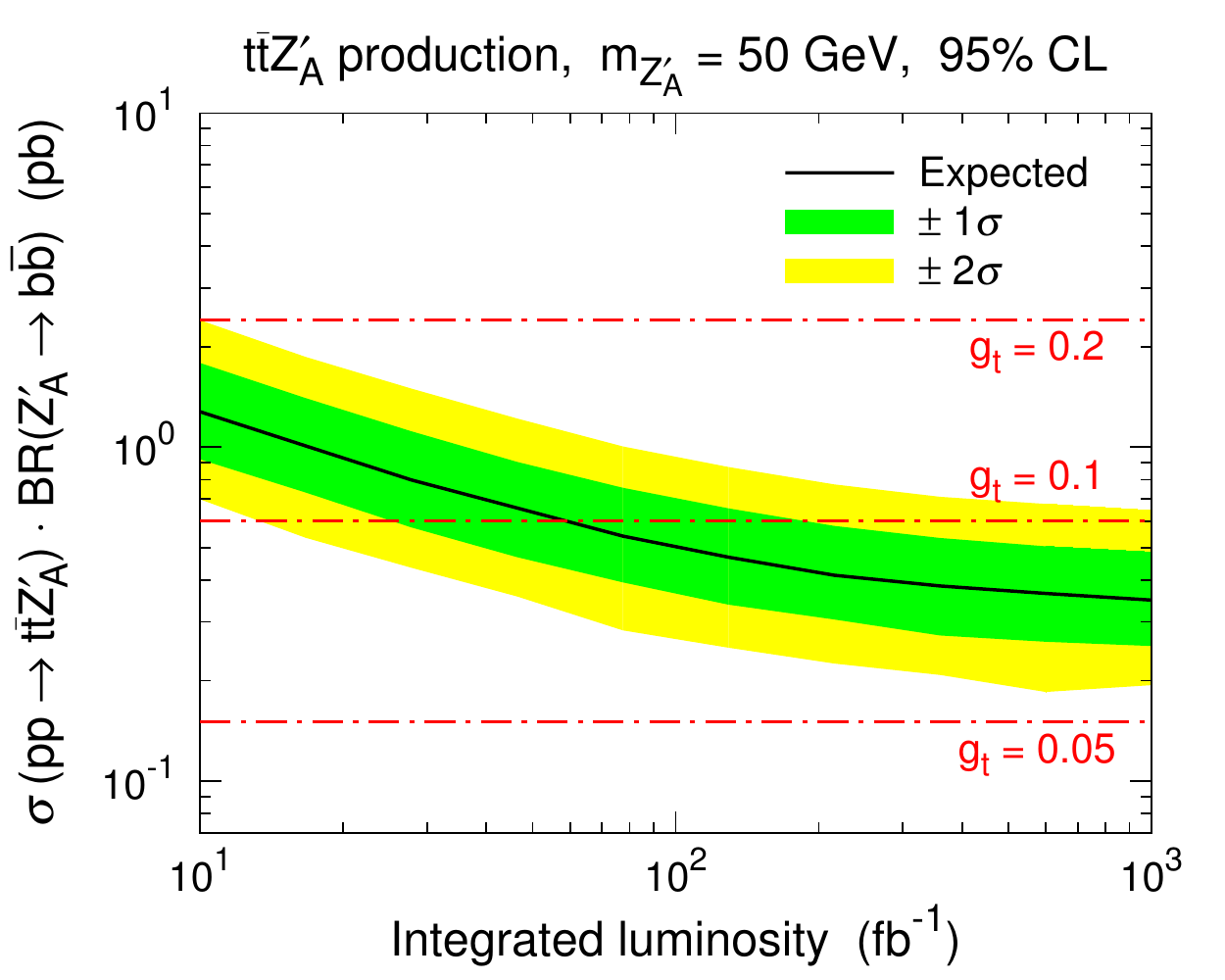}}
  \caption{This figure shows the expected 95\% CL exclusion limits on the signal strength $\sigma(pp\to\ttX) \cdot \BR(X\to b\bar{b})$ with $m_X = 50~\si{GeV}$ as functions of the integrated luminosity at the 14~TeV LHC for $\ttS$ (a), $\ttA$ (b), $\ttZpV$ (c), and $\ttZpA$ (d) production. The dot-dashed lines denote the signal strengths for the $g_t$ values labelled and $\BR(X\to b\bar{b}) = 100\%$.}
  \label{fig:CLs:lumi_Xsec}
\end{figure*}

To estimate the expected  exclusion on the signals we carry out a  $\CLs$ hypothesis test~\cite{Junk:1999kv,Read:2002hq} based on the $m_{bb}$ distributions from 15~GeV to 200~GeV shown in Fig.~\ref{fig:delphes:mbb}. We scale up the $\ttbb$ background by a factor of 1.2 in order to take into account the remaining backgrounds discussed earlier and assume a flat 10\% systematic uncertainty on the total background. The expected 95\% CL exclusion limits on the signal strength $\sigma(pp\to\ttX) \cdot \BR(X\to b\bar{b})$ as functions of the integrated luminosity are shown in in Fig.~\ref{fig:CLs:lumi_Xsec}.
These limits are comparable for the four simplified models due to the similarities in their production kinematics, and with the high-luminosity LHC it should be possible to bound the cross sections to the level of a few hundred femtobarns. For the pseudoscalar this corresponds to $g_t$ just under 1 (\textit{i.e.} essentially no suppression with respect to the SM Yukawa) while for the axial vector we can constrain $g_t$ down to 0.08.

\subsection{Expected Sensitivity for Discrimination among Simplified Models}

\begin{figure*}[t!]
  \centering
  \subfigure[~Normalised $m_{tt}$ distributions]{\includegraphics[width=0.46\textwidth]{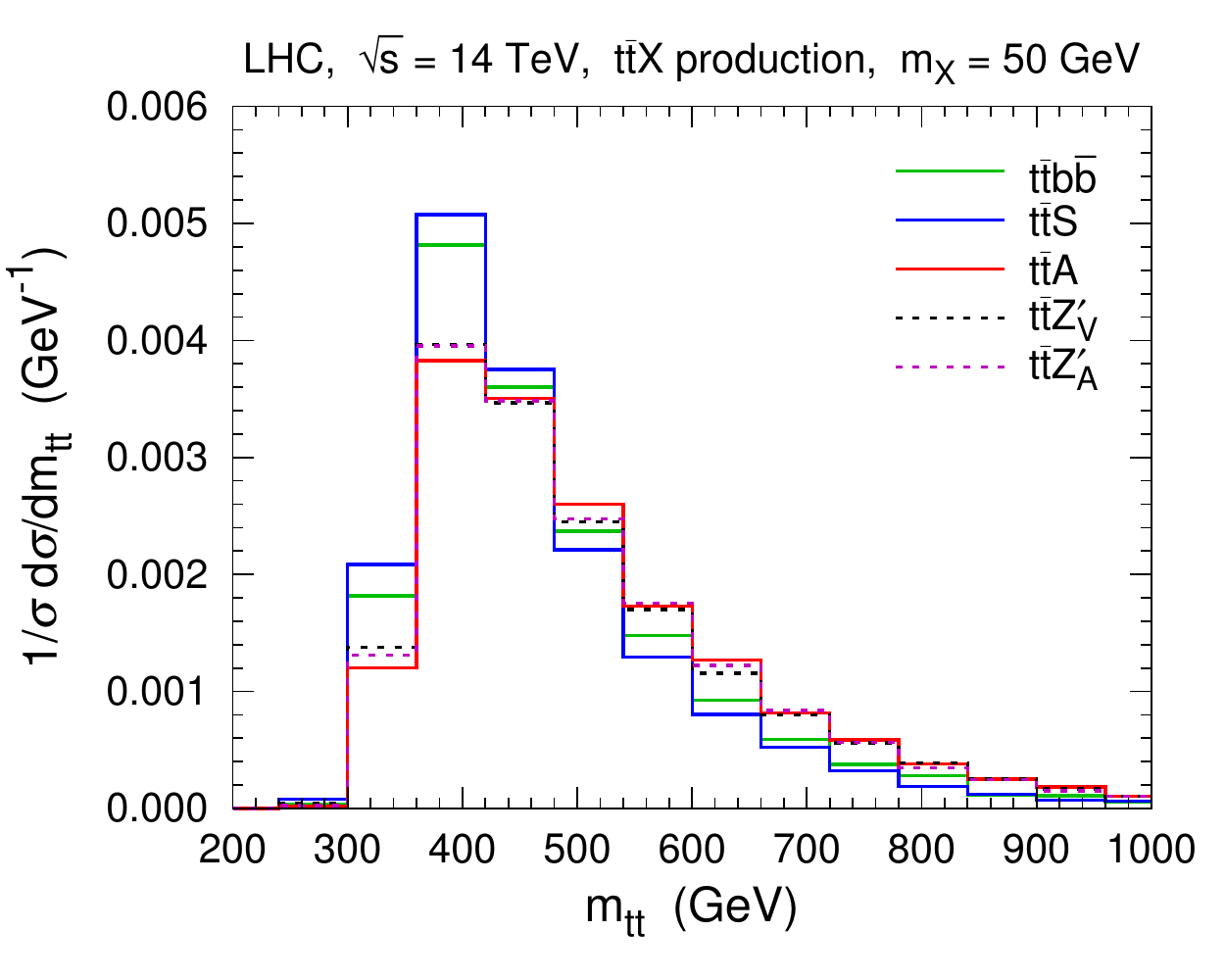}}
  \hfill
  \subfigure[~Normalised $\pTX$ distributions]{\includegraphics[width=0.46\textwidth]{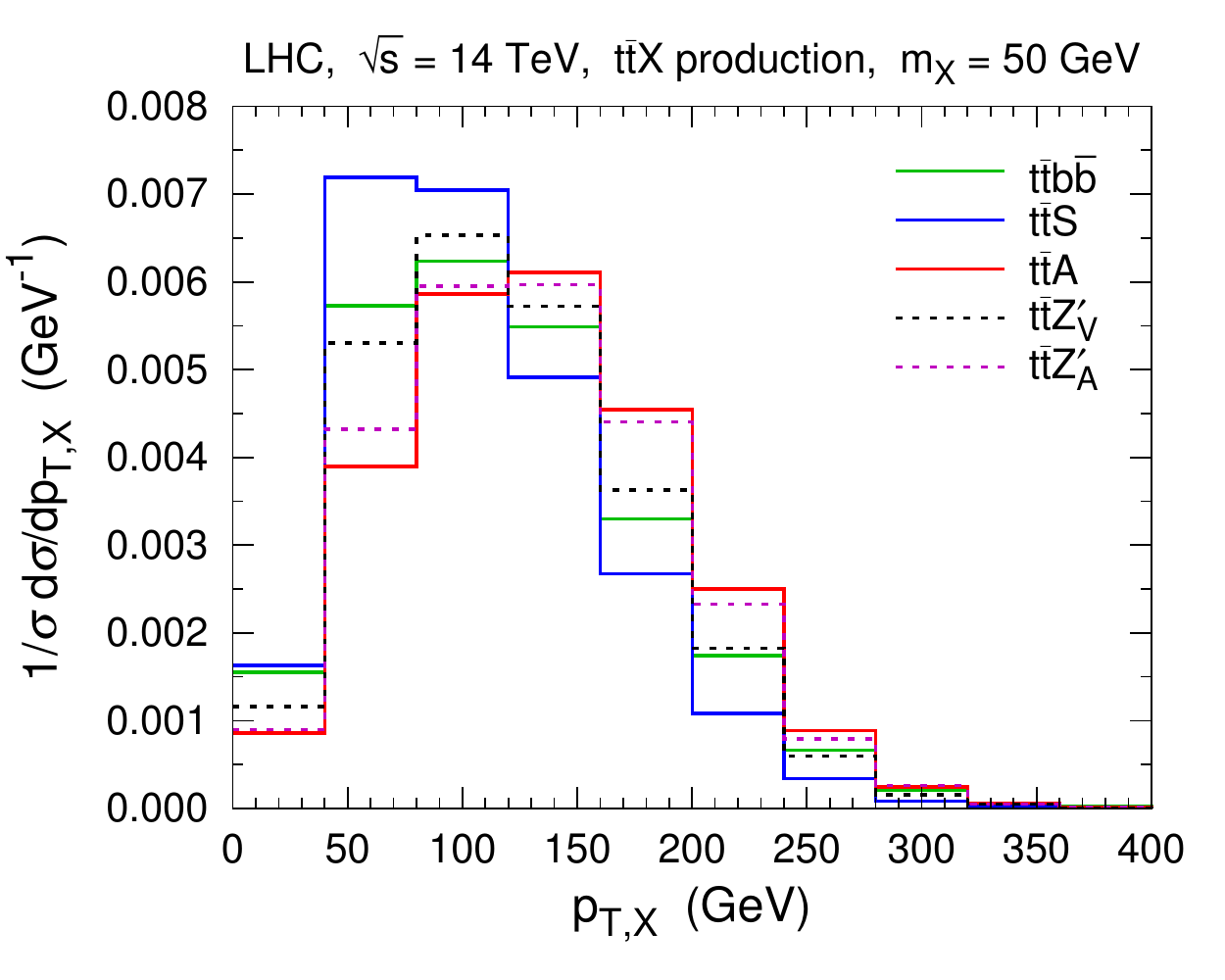}}
  \subfigure[~Normalised $\theta_{t,\mathrm{had}}^\mathrm{CM}$ distributions]{\includegraphics[width=0.46\textwidth]{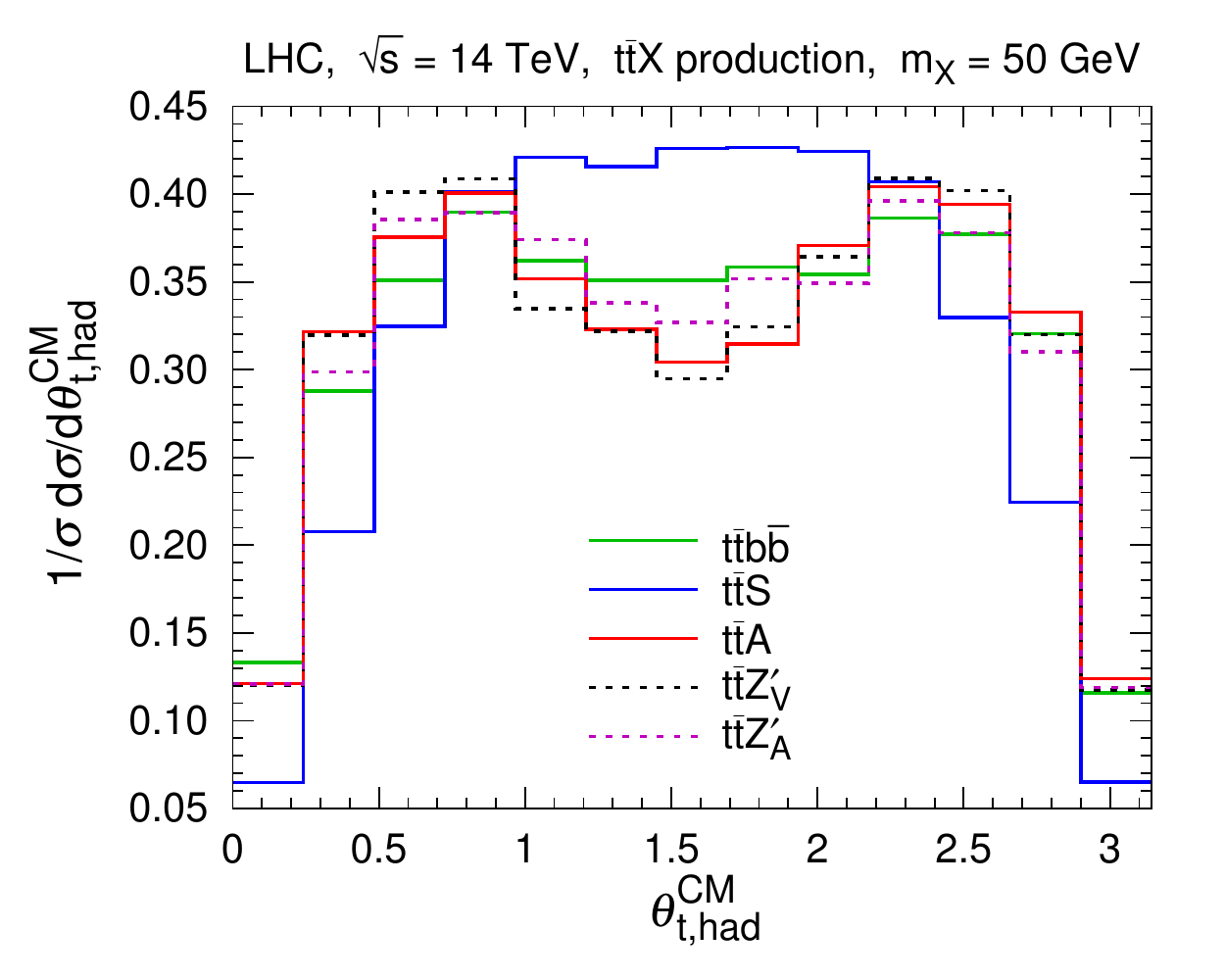}}
  \hfill
  \subfigure[~Normalised $\Theta^\mathrm{CM}$ distributions]{\includegraphics[width=0.46\textwidth]{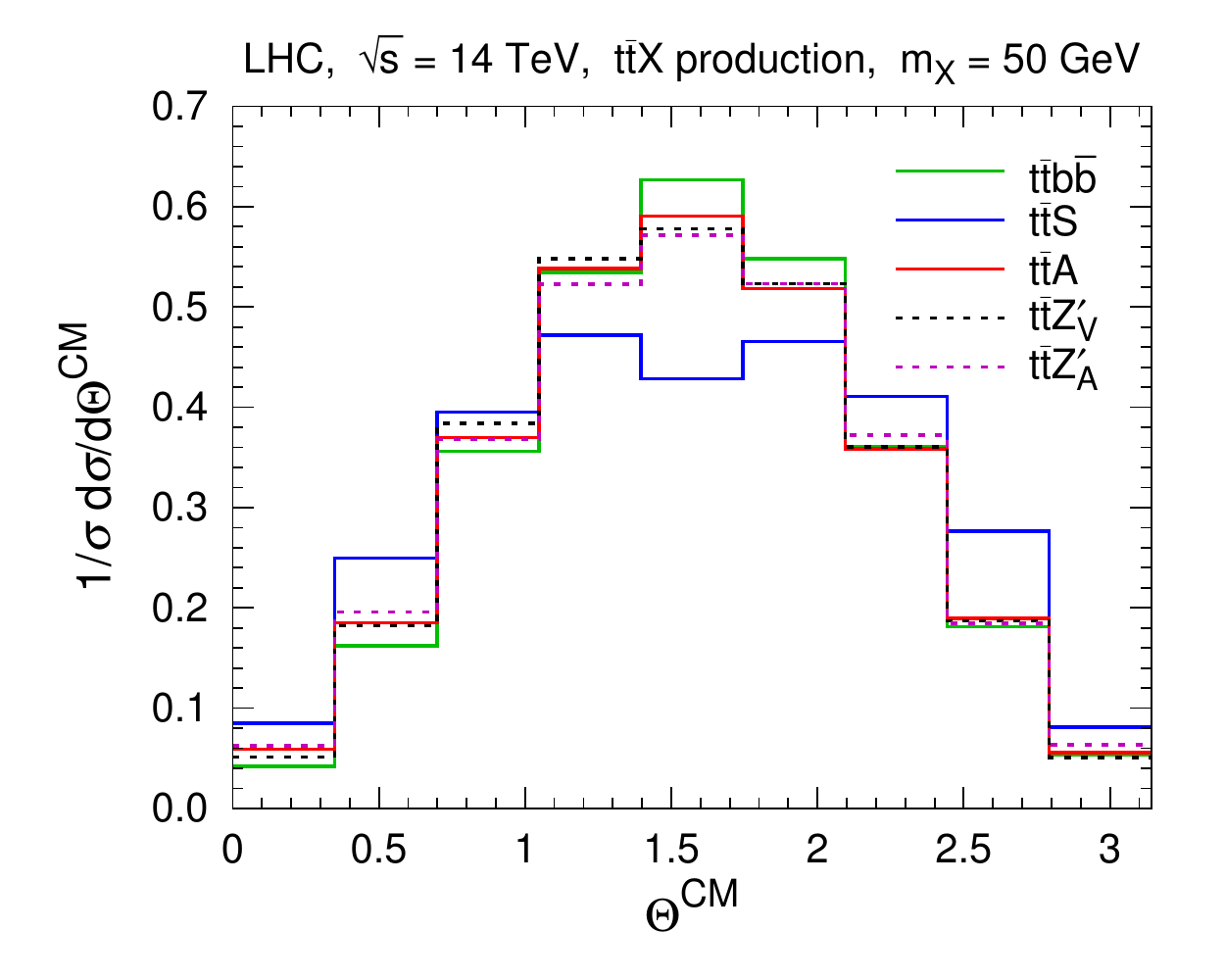}}
  \caption{This figure shows the normalised distributions of $m_{tt}$ (a), $\pTX$ (b), $\theta_{t,\mathrm{had}}^\mathrm{CM}$ (c), and $\Theta^\mathrm{CM}$ (d) at detector level for the 14~TeV LHC and $m_X=50~\si{GeV}$.}
  \label{fig:delphes:dist}
\end{figure*}

Through the above reconstruction procedure, we can construct the 4-momenta of the hadronically decaying top, the leptonically decaying top, and the resonance $X$ from the identified jets and lepton:
\begin{eqnarray}
p_{t,\mathrm{had}} &=& p_{b_1} + p_{j_1} + p_{j_2},\\
p_{t,\mathrm{lep}} &=& p_{b_2} + p_\ell + \slashed{p}_\mathrm{T},\\
p_X &=& p_{b_3} + p_{b_4}.
\end{eqnarray}
The missing momentum $\slashed{p}_\mathrm{T}$ only contains transverse components and hence the reconstructed $p_{t,\mathrm{lep}}$ is not as accurate as $p_{t,\mathrm{had}}$. We can find a CM frame where $\mathbf{p}_{t,\mathrm{had}} + \mathbf{p}_{t,\mathrm{lep}} + \mathbf{p}_X = 0$. Therefore, these 4-momenta allow us to construct discriminating variables $m_{tt}$, $\pTX$, $\theta_{t,\mathrm{had}}^\mathrm{CM}$, and  $\Theta^\mathrm{CM}$ that equivalent to the parton-level variables discussed in Sec.~\ref{sec:parton}. The normalised distributions for the signals and the $\ttbb$ background with all the selection cuts applied are shown in Fig.~\ref{fig:delphes:dist}. As expected, these detector-level variables catch the basic features of their parton-level counterparts demonstrated in Fig.~\ref{fig:parton:dist}. Note that $m_{tt}=(p_{t,\mathrm{had}}+p_{t,\mathrm{lep}})^2$ and $\theta_{t,\mathrm{had}}^\mathrm{CM}$ corresponds to the hadronically decaying top. An analogous variable $\theta_{t,\mathrm{lep}}^\mathrm{CM}$ can also be constructed from $p_{t,\mathrm{lep}}$, but it is less powerful than $\theta_{t,\mathrm{had}}^\mathrm{CM}$ for discrimination among simplified models.

We perform a $\CLs$ hypothesis test to investigate the discriminating power of each variables. Analogous to those in the ATLAS~\cite{Aad:2015mxa} and CMS~\cite{Khachatryan:2014kca} analyses for determining the spin and parity of the SM Higgs, the test statistic is defined as
\begin{equation}
Q=-2\ln\frac{\mathcal{L}(s_2+b)}{\mathcal{L}(s_1+b)},
\end{equation}
where $\mathcal{L}(s+b)$ denotes the likelihood for the background $b$ plus a signal hypothesis $s$. Thus, $Q$ is used to discriminate between signal hypotheses $s_1$ and $s_2$. For an observed value $Q_\mathrm{obs}$, the exclusion of the hypothesis $s_2$ in favour of the hypothesis $s_1$ (denoted as ``$s_1$ vs $s_2$'' below) is evaluated in terms of the modified confidence level
\begin{equation}
\CLs = \frac{P(Q\geq Q_\mathrm{obs}|s_2+b)}{P(Q\geq  Q_\mathrm{obs}|s_1+b)},
\end{equation}
where $P(Q\geq Q_\mathrm{obs}|s+b)$ is the probability for $Q\geq Q_\mathrm{obs}$ under a hypothesis $s$.

\begin{figure*}[t!]
  \centering
  \subfigure[~$m_{tt}$ variable]{\includegraphics[width=0.48\textwidth]{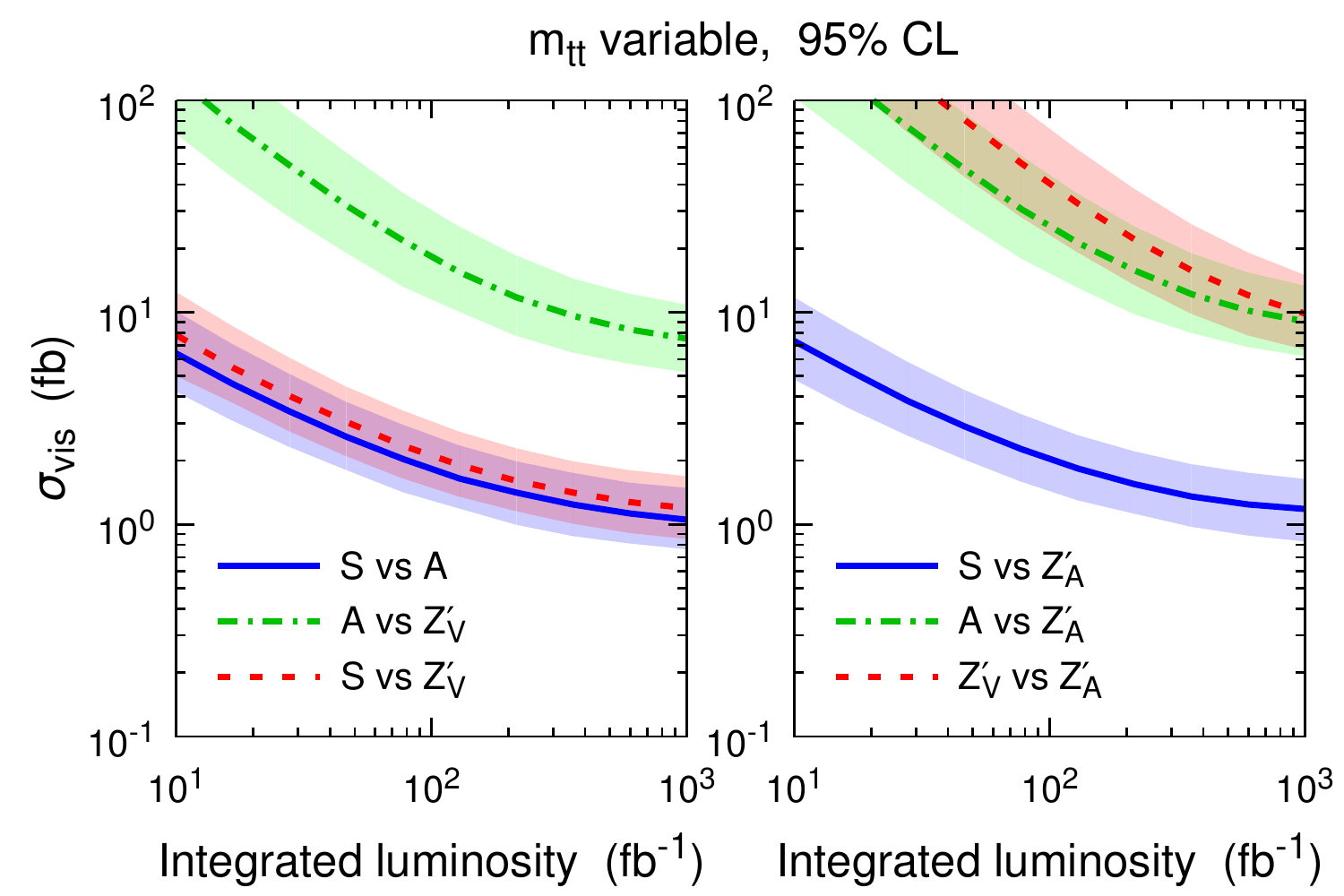}}
  \hfill
  \subfigure[~$\pTX$ variable]{\includegraphics[width=0.48\textwidth]{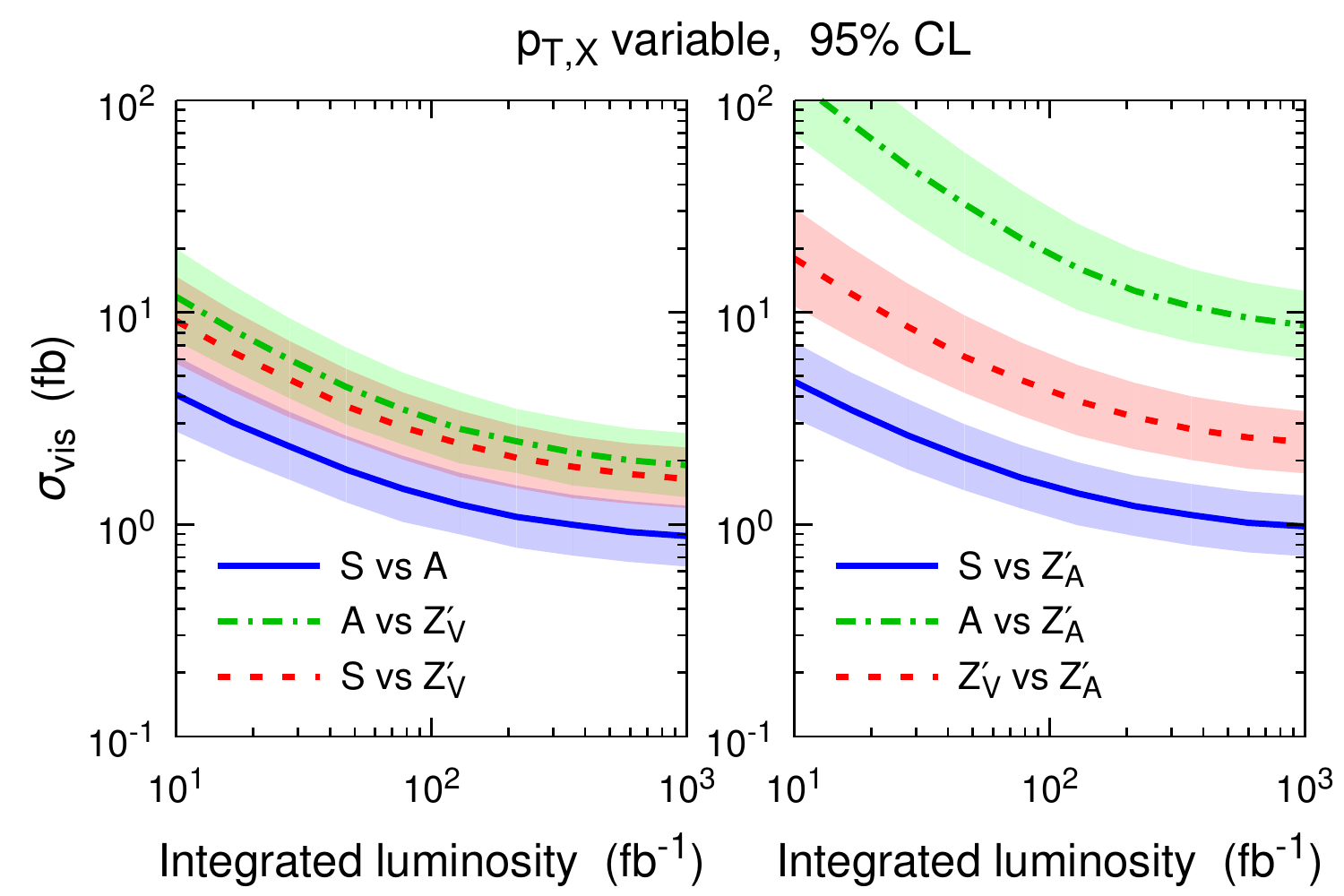}}
  \subfigure[~$\theta_{t,\mathrm{had}}^\mathrm{CM}$ variable]{\includegraphics[width=0.48\textwidth]{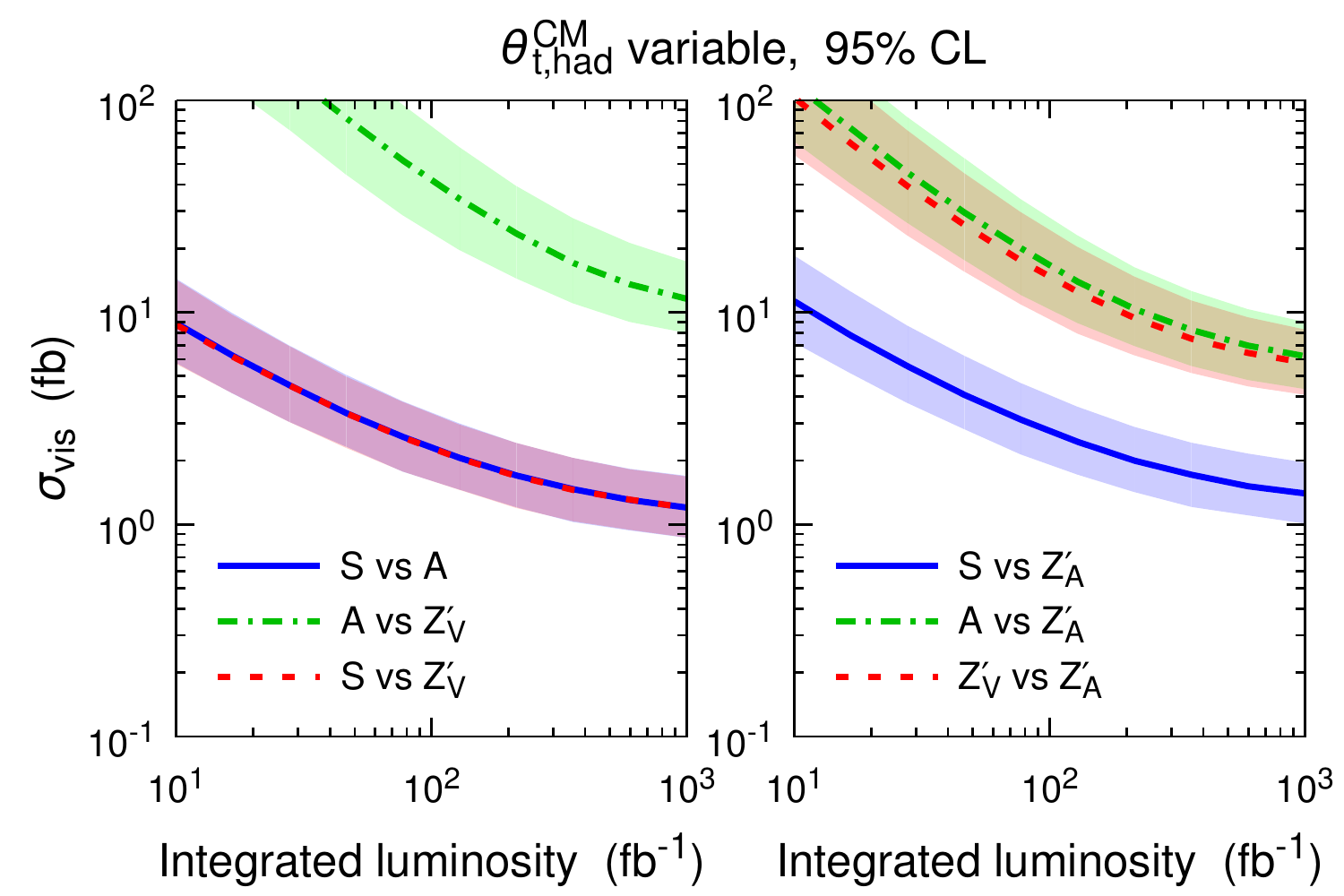}}
  \hfill
  \subfigure[~$\Theta^\mathrm{CM}$ variable]{\includegraphics[width=0.48\textwidth]{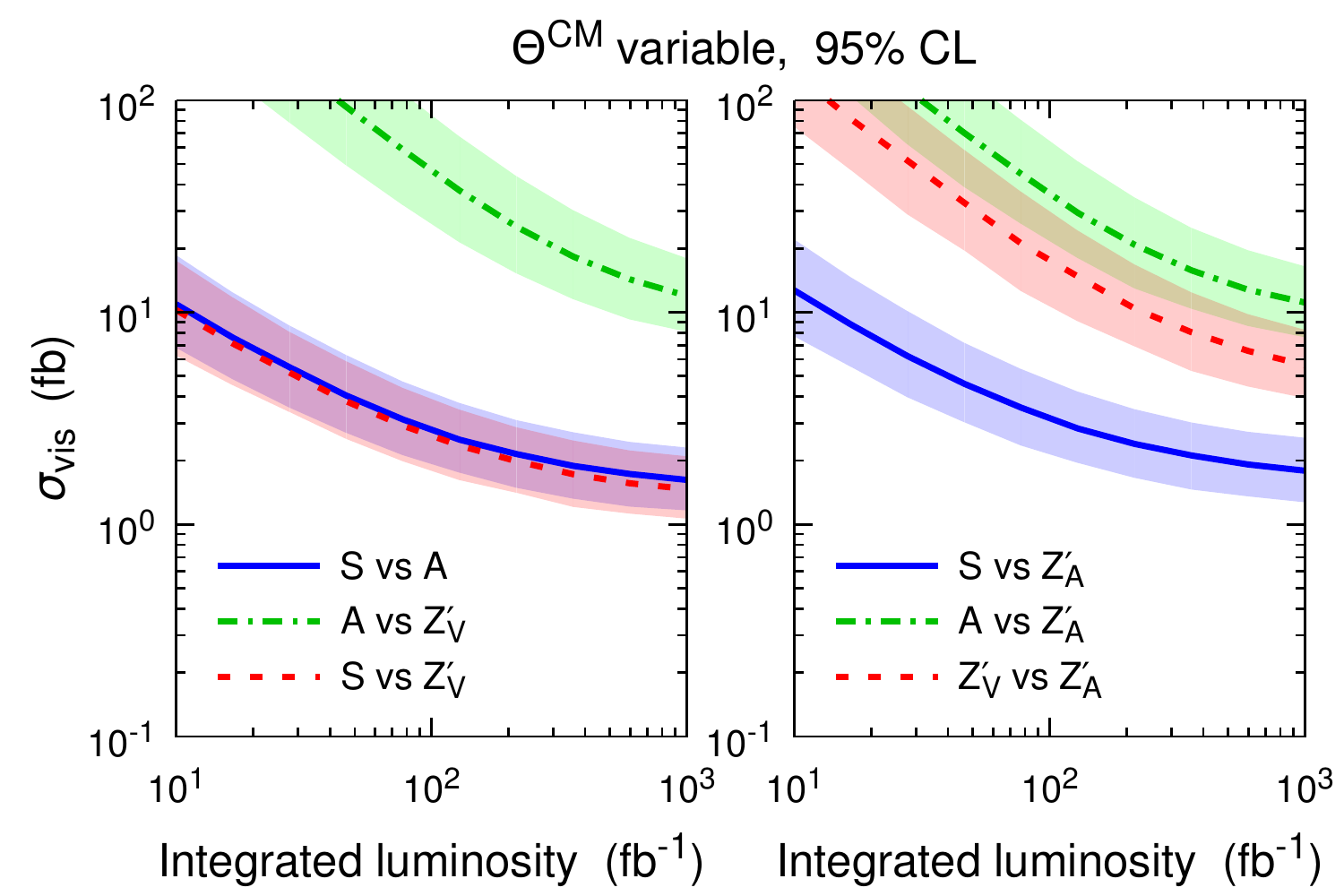}}
  \caption{This figure shows the expected 95\% CL exclusion limits on the visible cross section $\sigma_\mathrm{vis}$ as functions of the integrated luminosity at the 14~TeV LHC based on the variables $m_{tt}$ (a), $\pTX$ (b), $\theta_{t,\mathrm{had}}^\mathrm{CM}$ (c), and $\Theta^\mathrm{CM}$ (d). The lines denote the median value of the limit, while the coloured bands denote the $\pm 1\sigma$ range. ``$s_1$ vs $s_2$'' corresponds to the exclusion of the signal hypothesis $s_2$ in favour of the signal hypothesis $s_1$, assuming both hypotheses yield the same visible cross section.}
  \label{fig:discrim:lumi_visXsec}
\end{figure*}

Fig.~\ref{fig:discrim:lumi_visXsec} shows the expected 95\% CL exclusion limits on the visible cross section $\sigma_\mathrm{vis}$ as functions of the integrated luminosity based on the discriminating variables, assuming 10\% systematic uncertainty on the background. Here $\sigma_\mathrm{vis}$ is defined as the cross section taking into account the cut acceptance and efficiency. We assume each pair of signal hypotheses yield the same $\sigma_\mathrm{vis}$, and evaluate the exclusion limit of one hypothesis in favour of the other one. In this way, the differences among these limits only come from the different behaviours of the signal hypotheses in the distributions shown in Fig.~\ref{fig:delphes:dist}. Overall, the $\pTX$ variable seems to be the most powerful one, except for the ``$A$ vs $\ZpA$'' case, where the $\theta_{t,\mathrm{had}}^\mathrm{CM}$ variable is better than $\pTX$ for a high integrated luminosity of $\sim 1~\si{ab^{-1}}$. The $\ttS$ production is the easiest to be distinguished from the rest, because its distributions of all the discriminating variables behave most differently from others. The worst case is the discrimination between $A$ and $\ZpA$, which yield similar shapes for every variable.

\begin{figure*}[t!]
  \centering
  \subfigure[~$m_{tt}$ variable]{\includegraphics[width=0.48\textwidth]{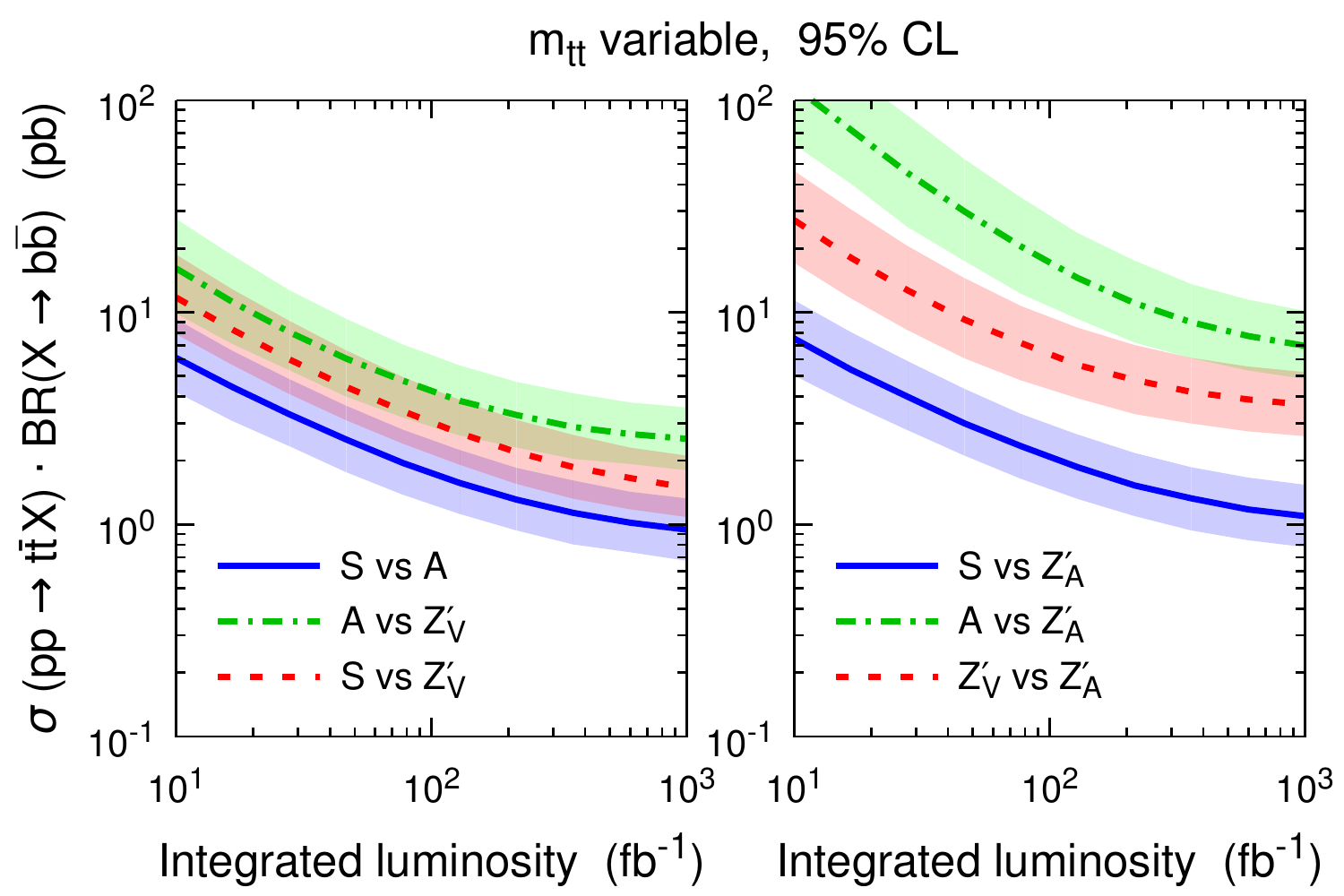}}
  \hfill
  \subfigure[~$\pTX$ variable]{\includegraphics[width=0.48\textwidth]{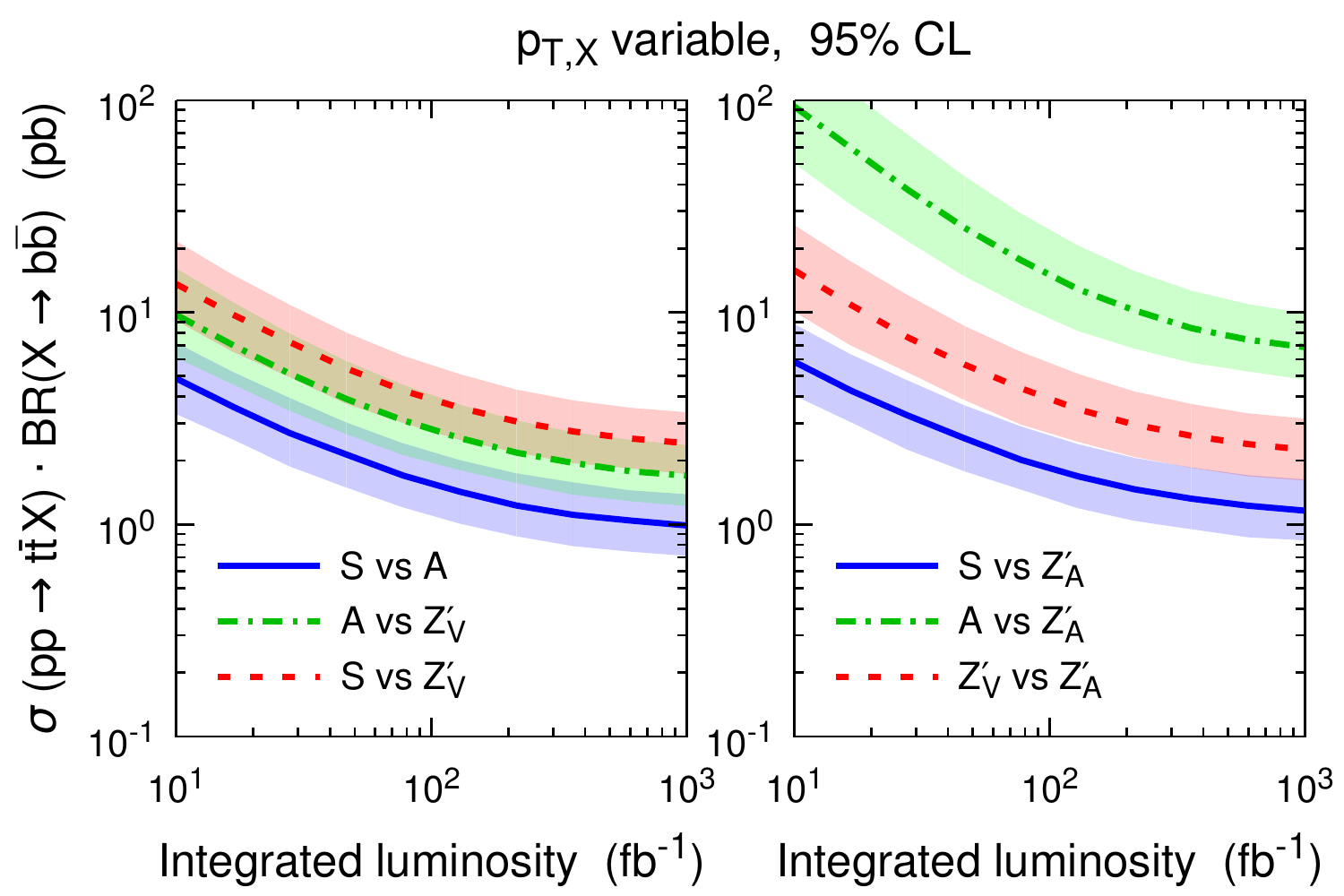}}
  \subfigure[~$\theta_{t,\mathrm{had}}^\mathrm{CM}$ variable]{\includegraphics[width=0.48\textwidth]{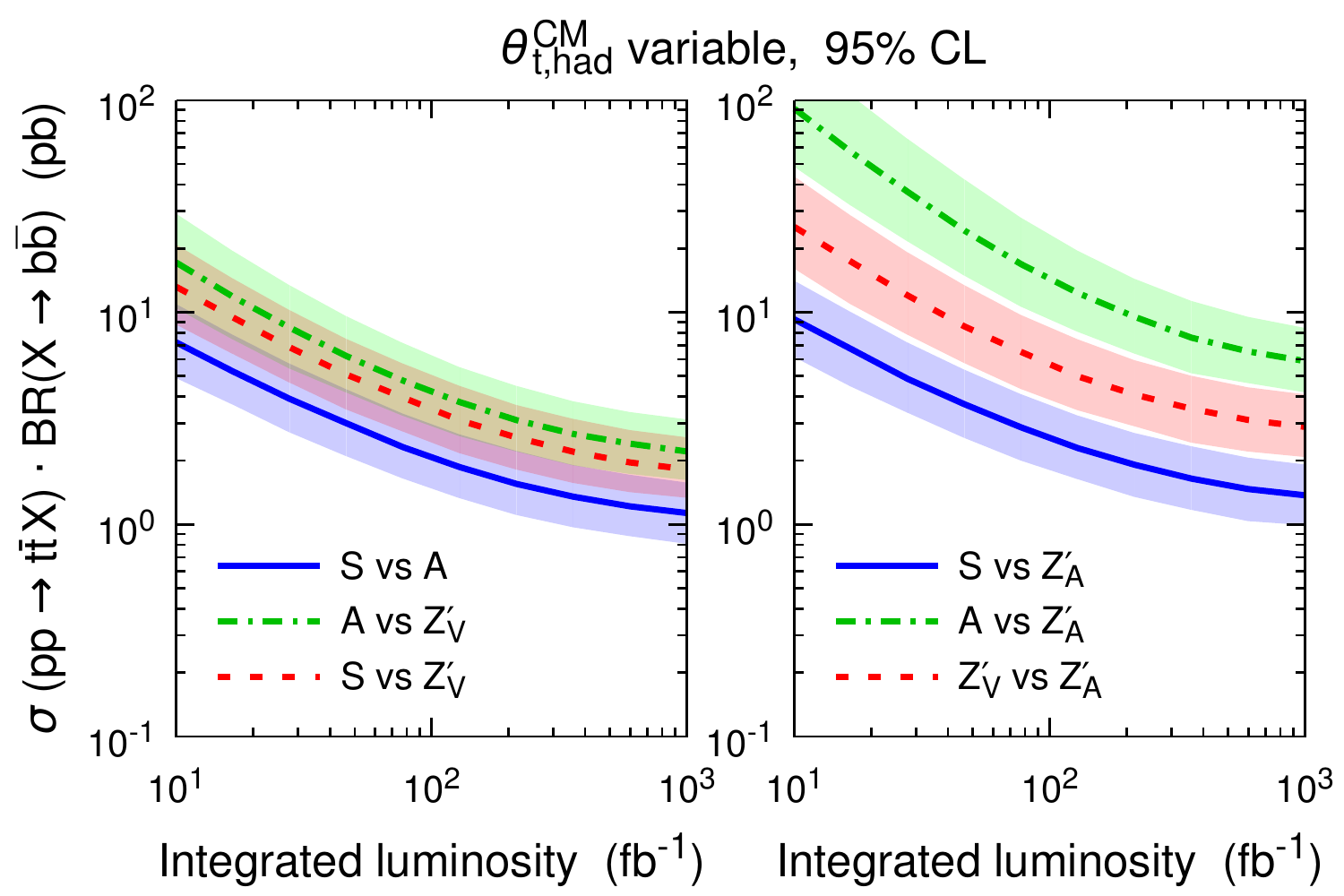}}
  \hfill
  \subfigure[~$\Theta^\mathrm{CM}$ variable]{\includegraphics[width=0.48\textwidth]{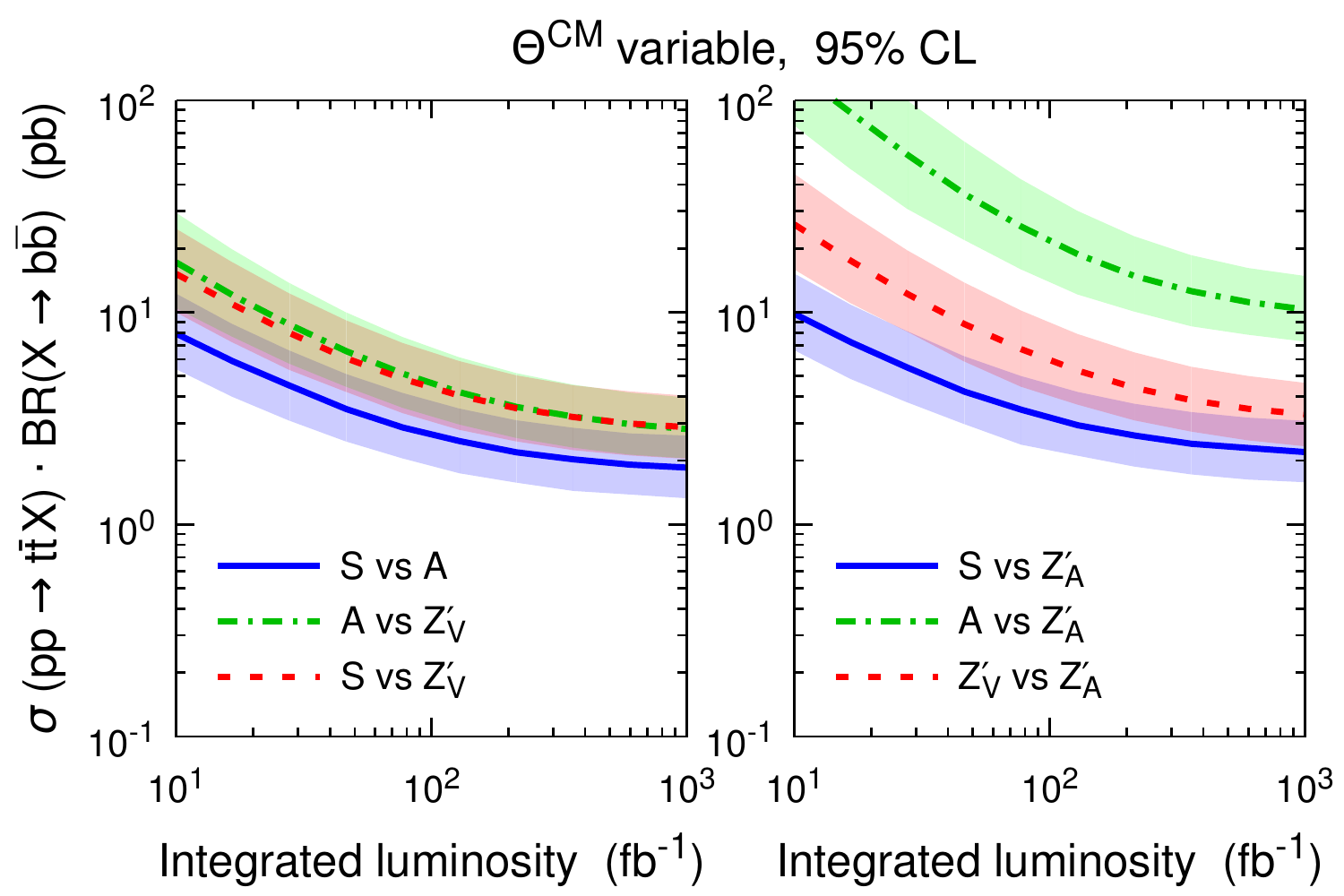}}
  \caption{This figure shows the expected 95\% CL exclusion limits on the signal strength $\sigma(pp\to\ttX) \cdot \BR(X\to b\bar{b})$ as functions of the integrated luminosity at the 14~TeV LHC based on the variables $m_{tt}$ (a), $\pTX$ (b), $\theta_{t,\mathrm{had}}^\mathrm{CM}$ (c), and $\Theta^\mathrm{CM}$ (d). The lines denote the median value of the limit, while the coloured bands denote the $\pm 1\sigma$ range. ``$s_1$ vs $s_2$'' corresponds to the exclusion of the signal hypothesis $s_2$ in favour of the signal hypothesis $s_1$, assuming both hypotheses yield the same signal strength.}
  \label{fig:discrim:lumi_Xsec}
\end{figure*}

On the other hand, we also assume two signal hypotheses yield the same signal strength $\sigma(pp\to\ttX) \cdot \BR(X\to b\bar{b})$ and estimate the corresponding exclusion limits. The results are presented in Fig.~\ref{fig:discrim:lumi_Xsec}. In this case, the selection efficiency of the cuts affects the exclusion limits. From Table~\ref{tab:cutflow}, we know that the cut efficiencies for the $\ttS$, $\ttA$, $\ttZpV$, and $\ttZpA$ production are \num{5.5e-4}, \num{7.8e-4}, \num{6.3e-4}, and \num{7.4e-4}, respectively for $m_X=50$~GeV.
Due to the higher cut efficiencies for $\ttA$ and $\ttZpA$ production than that of $\ttZpV$ production, the exclusion limits for ``$A$ vs $\ZpV$'' and ``$\ZpV$ vs $\ZpA$'' are greatly improved, compared with those in Fig.~\ref{fig:discrim:lumi_visXsec}. Given an integrated luminosity of $1~\si{ab^{-1}}$ and  a signal strength of $\sim 1-2~\si{pb}$, we can discriminate between any pair of simplified models at 95\% CL except for the worst-case scenario of ``$A$ vs $\ZpA$''.

\section{Summary and Conclusions}

Searches for $\ttX$ production at the LHC are sensitive to a new resonance $X$ coupled to the third generation quarks. If $X$ is discovered, a further measurement of its parity and spin will be essential for revealing the underlying new physics. In this work we assumed a class of simplified models to describe the couplings between $X$ and the third generation quarks, with $X$ being a scalar, pseudoscalar, vector, or axial vector. Then we sought kinematic variables that are helpful for determining parity and spin quantum numbers and investigated the expected sensitivity through detailed simulation.

We have demonstrated four parton-level variables which exhibit different shapes for different models. Two of them are defined in the $\ttX$ CM frame. Therefore, using them  requires a nearly full reconstruction of two tops and the resonance $X$, which can be achieved in the semi-leptonic channel. We have carried out the reconstruction procedure based on simulation in this channel and estimated the LHC sensitivity for discovery.

We constructed the detector-level counterparts of the parton-level variables and observed that their distributions preserve the important features for discrimination between the different simplified models. A $\CLs$ hypothesis test has been performed to evaluate the sensitivity of discrimination separately based on each variables. We found that the scalar is the easiest one to be distinguished from others while the hardest case is to discriminate between the pseudoscalar and the axial vector.

Further improvements to our analysis could be made by utilising jet substructure techniques to suppress the background more, and to allow a more accurate attribution of the $b$-jets used in the top and $X$ reconstruction. It would also be interesting to perform a combined analysis of leptonic and semi-leptonic final states to see the ultimate sensitivity of the LHC. We leave this for future work.

{\emph{Acknowledgements.}} 
MJD and ZHY are supported by the Australian Research Council. MS is supported in part by the European Commission through the ``HiggsTools'' Inital Training Network PITN-GA-2012-316704.

\bibliography{ttHA}

\end{document}